\documentclass{jfm}
\usepackage{url}
\usepackage[utf8]{inputenc}
\usepackage{amsfonts,amsmath,amssymb,graphicx}
\usepackage{subfig}
\usepackage{natbib}
\usepackage{bbm}
\usepackage{booktabs}
\newcommand{\uu}{\mathbf{u}}

\newcommand{\nab}{\boldsymbol{\nabla}}

\newcommand{\TT}{\mathbb{T}}

\newcommand{\RR}{\mathbb{R}}
\newcommand{\ZZ}{\mathbb{Z}}

\newcommand{\xx}{\mathbf{x}}

\newcommand{\Rey}{\mathrm{Re}}

\title[Dissipation by unsteady boundary layers]{
Energy dissipation caused\\ by boundary layer instability \\ at vanishing viscosity
}

\author[R. Nguyen van yen, M. Waidmann, R. Klein, M. Farge \and\  K. Schneider]{Romain 
Nguyen van yen$^1$, Mathias Waidmann$^1$, Rupert Klein$^1$, Marie Farge$^
2$ \and\  Kai Schneider$^3$}
\affiliation{
$^1$ Institut f\"ur Mathematik, Freie Universit\"at Berlin, Arnimallee 6,
 14185 Berlin, Germany \\
$^2$ LMD-CNRS, Ecole Normale Sup\'erieure, 24 rue Lhomond, 75231 Paris Cedex 5, France \\
$^3$ Institut de Math\'ematiques de Marseille, Aix-Marseille Universit\'e \& CNRS, Marseille, France
}

\begin{document}
\maketitle

\begin{abstract}
A qualitative explanation for the scaling of energy dissipation by high 
Reynolds number fluid flows in contact with solid obstacles is proposed
in the light of recent mathematical and numerical results.
%
%
Asymptotic analysis suggests that it is governed by a fast, small scale Rayleigh--Tollmien--Schlichting instability
with an unstable range whose lower and upper bounds scale as $\Rey^{3/8}$ and $\Rey^{1/2}$, respectively.
By linear superposition the unstable modes induce a boundary vorticity flux of order $\Rey^1$,
a key ingredient in detachment and drag generation according to a theorem of Kato.
These predictions are confirmed by numerically solving the Navier--Stokes 
equations in a two-dimensional periodic channel
discretized using compact finite differences in the wall-normal direction, 
and a spectral scheme in the wall-parallel direction.
\end{abstract}


\tableofcontents

\section{Introduction}

Since the challenge laid down by Euler in 1748 for the Mathematics Prize of the Prussian Academy of Sciences in Berlin,
the force exerted onto a solid by a fluid flow is one of the central unknowns in the field of fluid mechanics.
From the vorticity transport equations he had derived,
d'Alembert \citeyearpar{Rond1768} deduced his now famous paradox, that this force should vanish,
contrary to what experimental results and even everyday observation indicate.
A frictional explanation involving the viscosity of the fluid was advanced during the 19th century within the frame of the new theories of Navier, Saint-Venant, and Stokes, 
but the actual amplitude of the force remained unaccounted for.
Indeed, estimates based on the magnitude of the viscosity $\nu$ of the fluid, or equivalently, after non-dimensionalization, on the inverse of the Reynolds number $\Rey$,
predict that friction should become negligible when $\Rey \gg 1$.

The group working at G\"ottingen University, notably \cite{Prandtl1904} and \cite{Blasius1908},
were first to come up with a way of computing, under some hypotheses, the asymptotic behavior of the force in the limit $\Rey \to \infty$.
Their method relies on the notion that viscous effects are confined
to a layer of thickness of order $\Rey^{-1/2}$ along the boundary, now called Prandtl boundary layer (BL),
inside which the motion is governed by appropriately rescaled equations, 
whereas the bulk of the fluid remains inviscid.
Momentum transport across this layer gives rise to a net drag force of order $O(\Rey^{-1/2})$, 
which can even be computed explicitly in some academic cases.
Although Prandtl's theory has been very fruitful, 
it has the drawback of breaking down for sufficiently large $\Rey$ in practically all relevant flow configurations.
Indeed, according to the experiments made in G\"ottingen itself and elsewhere, the force then acquires the stronger scaling $O(1)$ \citep{Schlichting1979}.
%

The precise dynamical mechanism which allows a transition to this regime,
starting from a fluid initially at rest, is still unknown today.
Prandtl, assuming that the BL approximation becomes invalid beyond a certain separation point along the boundary,
had already established that the viscous shear vanishes or, in other words,
that the flow in the neighborhood of the wall reverses direction at this point.
After the formal asymptotic expansion achieved by \cite{Goldstein1948},
which clearly indicates a singularity,
a viscous scaling of the parallel coordinate was developed to analyze the flow near the separation point,
finally leading to the so-called triple deck structure \citep{Stewartson1974}.
This steady asymptotic theory is instructive and well developed but remains unsatisfactory, 
since in practice all flows are unsteady above a certain Reynolds number.

The classical understanding of the onset of unsteadiness in shear flows goes back to \cite{Rayleigh1880},
who showed in particular that a necessary condition for inviscid instability is that the base velocity profile
has an inflexion point. 
Since many BL flows lack such a point,
Rayleigh's result seemed at first to rule out linear instability as a generic mechanism for BL breakdown.
This impression was even reinforced by a result of \cite{Sommerfeld1909} showing that purely  linear shear flows ($u''(y)=0$)
were linearly unconditionally stable even when including the effect of viscosity.
But the G\"ottingen group then realized that including the effects of viscosity
with a nonzero second derivative of the base flow could trigger an instability which is absent in the inviscid setting
\citep{Prandtl1921,Tietjens1925}.
Following some of the ideas developed by \cite{Heisenberg1924} in his thesis,
\cite{Tollmien1929} then studied asymptotic solutions to the Orr--Sommerfeld equation,
and thus achieved an elegant analysis of the instability mechanism.
He also produced an approximate marginal curve
for what is now known as the Tollmien--Schlichting instability,
a widely accepted mechanism for transition to turbulence in BLs.
Its range of application is however limited to small perturbations around a stationary base flow.

Beyond this, one is faced with the full unsteady Navier--Stokes equations (NSE) at high Reynolds number,
and the BL problem becomes so wide reaching that it has been investigated almost independently 
for the past 60 years by two disctinct schools, 
which we will call the `aerodynamical' and the `mathematical' schools.
Aerodynamicists took steady BL theory as their starting point
and attempted to generalize the main ideas of Prandtl and Goldstein to the unsteady case.
This lead in the 1950s to the establishment of the Moore-Rott-Sears criterion \citep{Sears1975}, 
stating that detachment originates from a point within the BL, not necessarily lying on the wall,
where vorticity vanishes and parallel velocity equals that of the exterior flow.
Reasoning by analogy with the steady case, \cite{Sears1975} conjectured that separation coincides with the appearance of a singularity in the solution at some finite time $t^*$.
The approach to such a singularity was confirmed using numerical experiments by \cite{VanDommelen1980} for the impulsively started cylinder, 
and then analyzed in detail by power series expansions in Lagrangian variables by \cite{VanDommelen1990}.
We refer the reader to van Dommelen's PhD thesis \citeyearpar{VanDommelen1981}
for a detailed pedagogical account of these findings.
Later work largely supports his results \citep{Ingham1984,E1997,Gargano2009}.


The next challenge was to understand what happens to the actual NSE flow as the corresponding Prandtl solution becomes singular.
\cite{Elliott1983} obtained the estimate $(t^*-t) = O(\Rey^{-2/11})$
for the time $t$ at which the BL assumptions first break down.
To describe the solution at later times, 
some hope came from the interacting boundary layer (IBL) method, which relieves the Goldstein singularity in steady BLs, 
but it was quickly shown \citep{Smith1988,Peridier1991,Peridier1991a} to lead to a finite time singularity when applied to unsteady problems.

Over the same period of time, the mathematical school focused on totally different issues concerning
the initial value problem for the unsteady Prandtl equations on the one hand, and on the vanishing viscosity limit
for solutions of the NSE on the other hand.
Local well-posedness for the Prandtl equations was proved long ago by \cite{Oleinik1966}
in cases where detachment is not expected, {\it i.e.}, for monotonous initial data and favorable pressure gradient.
\cite{Sammartino1998ab} showed local well-posedness without the monotonicity conditions,
but this time under a very harsh regularity condition, that the amplitude of the parallel Fourier coefficients of all quantities decrease \textit{exponentially} with wavenumber.
These conditions were recently improved by \cite{Gerard-Varet2013}, but 
the required decay of Fourier coefficients is still faster than any power of $k$.
\cite{Maekawa2014}  proved the convergence of the Euler and Prandtl solution versus the Navier--Stokes solution in the $L^\infty$ norm with order $\sqrt{\nu}$
for an initial condition where the vorticity is compactly supported at finite distance from the wall.

\cite{Kato1984} made a decisive contribution, by linking the vanishing viscosity limit
problem to the behavior of energy dissipation at the boundary.
Kato's theorem implies, in the particular case of a flow in a smooth two-dimensional domain $\Omega$ with smooth initial data and without forcing, 
the equivalence between the following assertions:
\begin{enumerate}
\item the NS flow converges to the Euler flow uniformly in time in the energy norm,
\item  the energy dissipation associated to the NS flow, integrated over a strip of thickness proportional to $\Rey^{-1}$ around the solid,
which we will call the Kato layer, tends to zero as $\Rey \to \infty$. 
\end{enumerate}
Since convergence to the Euler flow excludes detachment, one of the essential messages carried by this theorem is that the flow has to develop dissipative activity at a scale at least as fine as $\Rey^{-1}$ for detachment to be possible. 
Later refinements of Kato's work linked breakdown to scalings $O(\Rey^\alpha)$ with $\alpha \geq 3/4$ of the wall pressure gradient \citep{Temam1997},
and $O(\Rey^{\beta}$ with $\beta \geq 1/2$ of the $L^2$ norm of velocity in the Kato layer \citep{Kelliher2007}.

Not only is the gap between the aerodynamical and the mathematical schools of thought
quite impressive when one realizes that they are really concerned with the same problem,
but after close consideration, it even appears that they contradict each other on an essential point
which we now attempt to clarify.
As shown by \cite{VanDommelen1990}, the finite time singularity in the unsteady Prandtl equations,
which is characterized by a blow-up of parallel vorticity gradients,
does correspond to a ``detachment'' process, 
in the sense that there exist fluid particles that are accelerated infinitely rapidly away from the wall.
In the following, we shall call this process ``eruption''.
Since its discovery, it seems to have been at least tacitly assumed
to underly the initial stage of the ({\it a priori} different) detachment process 
actually experienced by the NS solution.
According to this scenario, singularity would be avoided in the NS case
thanks to a large scale process not taken into account in the Prandtl approximation, namely the normal pressure gradient.
The Kato criterion, on the other hand, tells us something entirely different,
which is that for detachment to happen, 
scales as fine as $\Rey^{-1}$, which are not even accounted for in the Prandtl solution, need to come into play.
This suggests that eruption never really enters the scene in the NS flow, 
being indeed short-circuited by a faster mechanism at finer scales.

In the last decade, instabilities at high parallel wavenumber came up as a possible explanation for these finer scales.
On the mathematical side, \cite{Grenier2000} proved that Prandtl's asymptotic expansion is invalid for
some types of smooth perturbed shear flows, due to instabilities at high parallel wavenumbers.
Then \cite{Gerard-Varet2010} showed, again for smooth perturbed shear flows,
that the Prandtl equations could be linearly ill-posed in any
Sobolev space ({\it i.e.}, assuming spectra which decay like powers of $k$),
although they are locally well-posed in the analytical framework,
as we have seen above.
Once again, the details of the proof show that the ill-posedness is due to modes with large wavenumber in the direction parallel to the wall.
More recently, \cite{Grenier2014,Grenier2014a,Grenier2015} started working on the ambitious program aiming to achieve a rigorous mathematical description
of instabilities in generic BL flows.

Several fluid dynamicists also advanced the idea that
instability-type mechanisms may play an important role in the process of unsteady detachment.
\cite{Cassel2000} directly compared numerical solutions of the NSE and of the corresponding Prandtl equations 
in an attempt to verify the correctness of the asymptotic expansion.
Although he considered a vortex induced BL instead of the impulsively started cylinder, 
his Prandtl solution behaves qualitatively like the one in \cite{VanDommelen1980},
developing strong parallel velocity gradients in a process which seems to lead to a finite time singularity
associated to a large normal displacement of fluid particles concentrated around a single parallel location.
But interestingly, around the same time the NS solution adopts a quite different behavior,
characterized by the appearance of strong oscillations in the wall-parallel pressure gradient,
which he was not able to explain.
\cite{Brinckman2001} also saw oscillations, and claimed that they were due to a Rayleigh instability
of the shear velocity profile, an hypothesis which was developed further in the review paper by \cite{Cowley2002}.
Although the numerics underlying these findings,
and therefore their interpretation in terms of a Rayleigh instability, were later invalidated due to their insufficient grid resolution by \cite{Obabko2002},
the existence of an instability mechanism has continued to be a hot topic in later papers on unsteady detachment \citep{Bowles2003,Bowles2006,Cassel2010,Gargano2011,Gargano2014}.
In fact, it seems to be the only surviving conjecture at this time to explain unsteady detachment.
The nature and quantitative properties of the instability remain to be elucidated.

Imposing no-slip boundary conditions to high Reynolds number flows, even in 2D, is a tough numerical problem and
one should be especially careful with the numerics given that the problem is theoretically not well understood yet.
In our previous work \citep{Nguyenvanyen2011a}, 
we computed a series of dipole-wall collisions, a well studied academic flow introduced by \cite{Orlandi1990}.
Our goal was to derive the scaling of energy dissipation when the Reynolds number increases, for fixed initial data and geometry.
According to the Kato criterion, this is an important element to understand detachment.
We chose to work with a volume penalization scheme which has the advantages of being efficient,
easy to implement and most importantly to provide good control on numerical dissipation.
However, the no-slip condition is only approximated, and the higher the Reynolds number, the more costly it is to enforce satisfactorily.
In fact, post-processing numerically calculated flows revealed that they effectively experienced Navier boundary conditions with a slip length proportional to $\Rey^{-1}$
\citep[for a more detailed analysis of the scheme, see][]{Nguyenvanyen2014}.
In this setting, we did find indications that energy dissipation converges to a finite value when $\Rey \to \infty$.

\cite{Sutherland2013} then confirmed our findings using a Chebychev method with exact Navier boundary conditions,
but in the no-slip case concluded that energy dissipation vanishes, in accordance with an earlier claim of \cite{Clercx2002}.
This is quite unexpetected since, due to the spectral properties of the Stokes operator, the no-slip boundary condition is stiffer that any Navier boundary condition with a nonzero slip length, and should thus generate larger gradients or in other words \textit{more} dissipation.
In fact, looking more closely at their results, it appears that the central claim
is based on a single computation (see Figs.~17 and 18 of their paper),
for which a convergence test is not provided,
which in our opinion leaves the matter unsettled.

This is why we propose to revisit the issue once again, 
but this time using a numerical scheme which has both high precision and accurate no-slip boundary conditions.
For this, we have turned to compact finite differences with an \textit{ad hoc} irregular grid in the wall-normal direction, 
and Fourier coefficients in the wall-parallel direction.
Combining ideas developed in the last decade by many authors, we propose a heuristic scenario for detachment, 
based on an instability mechanism of the Tollmien--Schlichting type,
which also explains the new vorticity scaling $O(Re^1)$ and the occurence of dissipation.
We then check that all these processes are actually ocurring in our numerical solution of the dipole-wall initial value problem.
For this, adopting a methodology similar to the one employed by \cite{Cassel2000},
we proceed by direct comparison of the NS solution and of the corresponding Prandtl approximation.

In the first section, we introduce the flow configuration and the corresponding NS and Prandtl models.
Although these models are classical, we present specific reformulations which were chosen 
in order to facilitate both numerical efficiency and theoretical interpretation.
Then we use the model to predict the appearance of an instability
and understand its characteristics.
In the second section, we introduce the model discretizations which we have used for our numerical computations.
In the third section we describe the numerical results.
Finally, we analyze the numerical results in the light of our preceding theoretical developments and we draw the necessary conclusions.


\section{Models}

	\subsection{Navier--Stokes model}

The incompressible Navier--Stokes equations in a smooth plane domain $\Omega$ read
\begin{subequations}\label{eq:NSE}
\begin{align}
& \label{eq:momentum_transport} \partial_t \uu + \left(\uu \cdot \boldsymbol{\nabla} \right) \mathbf{u} = \nu \Delta \uu - \nab p \\
& \label{eq:NSE_div} \boldsymbol{\nabla} \cdot \mathbf{u} = 0,
\end{align}
where $\uu(\xx,t)$ is the velocity field, $p$ is the pressure field, $\nu$ is the kinematic viscosity,
and we shall denote $(u,v)$ the two components of $\uu$.
In order to make formulas a little more concise, we shall in the following often omit to write the time variable explicitly, except when doing so would create an ambiguity.

As spatial domain, we choose
\begin{equation}
\Omega = \left\{(x,y) \mid x \in \TT, y \in ]0,1[ \right\},
\end{equation} 
where $\TT = \RR/\ZZ$ is the unit circle,
which models a periodic channel.
Dirichlet boundary conditions are imposed at $y=0$ and $y=1$,
\begin{equation}
\label{eq:NSE_bc} \uu(x,0,t) = \uu(x,1,t) = 0,
\end{equation} 
and we specify an initial condition
\begin{equation}
\label{eq:NSE_ic} \uu(x,y,0) = \uu_i(x,y),
\end{equation} 
\end{subequations}
which we shall assume to have zero spatial average.
By introducing characteristic velocity and length scales $U$ and $L$, a Reynolds number can be defined as follows:
\begin{equation}\label{eq:def_reynolds}
\Rey = \frac{UL}{\nu}
\end{equation}

When discretizing this system, difficulties arise due to the interplay between 
the divergence condition (\ref{eq:NSE_div}) and the no-slip boundary condition (\ref{eq:NSE_bc}).
We have chosen to work with the vorticity formulation of the NSE,
which eliminates the divergence constraint at the cost of transforming the Dirichlet boundary conditions on $u$
into a non-local integral constraint on $\omega$.
Although they used to be controversial, such formulations are now well established \citep[see][]{Gresho1991,E1996,Maekawa2013},
under the condition that the discretization of the integral constraint is properly carried out.
Fortunately, our periodic channel geometry allows for an explicit and easy to understand approach which we now present.

%
The vorticity field $\omega = \partial_x v - \partial_y u$ satisfies the transport equation
\begin{equation}
\label{eq:vorticity_transport} 
\partial_t \omega + \boldsymbol{\nabla} \cdot (\mathbf{u} \omega) = \nu \Delta \omega,
\end{equation}
with initial data
\begin{equation}
\label{eq:vorticity_ic} 
\omega_i(x,y) := \nab \times \uu_i(x,y),
\end{equation}
where $\uu$ is expressed as a function of $\omega$ by means of the stream function $\psi$ defined by
\begin{subequations}\label{eq:psi_def}
\begin{align}
\label{eq:psi_def_x} u & = -\partial_y \psi \\  
\label{eq:psi_def_y} v & = \partial_x \psi,     
\end{align}\end{subequations}
which in turn satisfies the Poisson equation
\begin{equation}\label{eq:psi_eq}
\Delta \psi = \omega.
\end{equation}
From the wall-normal component of (\ref{eq:NSE_bc})
and the fact that $\int_\Omega \uu = 0$ is a constant of motion,
a Dirichlet boundary condition for $\psi$ follows,
\begin{equation}\label{eq:psi_bc}
\psi(x,0,t) = \psi(x,1,t) = 0,
\end{equation}
which uniquely determines $\psi$, and therefore $\uu$, as a function of $\omega$.

To close the problem, the tangential component of (\ref{eq:NSE_bc}), 
which has not yet been used, needs to be reformulated into the missing boundary condition on $\omega$
necessary because of the presence of a Laplacian in (\ref{eq:vorticity_transport}).
A general discussion of this issue has been carried out by \cite{Gresho1991}.
In our case, due to the simple geometry, (\ref{eq:psi_def})-(\ref{eq:psi_bc}) can be solved explicitly to get an expression for $\psi$.
For this, we first introduce the Fourier coefficients
\begin{equation}\label{eq:fourier_def}
\widehat{\omega}_k(y) = \int_0^1\omega(x,y) e^{-\iota \kappa x} \mathrm{d}x,
\end{equation}
where $\kappa = 2\pi k$, and the corresponding reconstruction formula
\begin{equation}\label{eq:fourier_reconstruction}
\omega(x,y) = \sum_{k\in\ZZ} \widehat{\omega}_k(y) e^{\iota \kappa x},
\end{equation}
which applies similarly for other fields.
By (\ref{eq:psi_eq}) we then have
\begin{equation}\label{eq:psi_eq_fourier}
-\kappa^2 \widehat{\psi}_k +  \partial_y^2 \widehat{\psi}_k = \widehat{\omega}_k.
\end{equation}
Combining this with the boundary conditions (\ref{eq:psi_bc}), we obtain
\begin{subequations}
\label{eq:psi_explicit}
\begin{multline}
\widehat{\psi}_k(y) = -\frac{1}{2\vert \kappa \vert \left(1-e^{-2\vert \kappa \vert}\right)} \Bigg( \left(1-e^{-2 \vert \kappa \vert y}  \right) \int_y^1 \widehat{\omega}_k(y')\left(e^{-\vert \kappa \vert(y'-y)} - e^{-\vert \kappa \vert(2-y'-y)}\right)\mathrm{d}y' \\ 
+ \left(1-e^{-2 \vert \kappa \vert (1-y)}  \right) \int_0^y \widehat{\omega}_k(y')\left(e^{-\vert \kappa \vert(y-y')} - e^{-\vert \kappa \vert(y+y')}\right)\mathrm{d}y' \Bigg),
\end{multline}
for $k \neq 0$, and
\begin{equation}
\widehat{\psi}_0(y) = y \int_0^1 (1-y') \widehat{\omega}_0(y') \mathrm{d}y' + \int_0^y (y-y') \widehat{\omega}_0(y') \mathrm{d}y'.
\end{equation}
\end{subequations}

Using these expressions, the no-slip boundary condition (\ref{eq:NSE_bc}) can now be reformulated as two linear constraints on $\widehat{\omega}_k$:
\begin{equation}\label{eq:integral_bc_old}
\forall k, B^+_k \left(\widehat{\omega}_k \right) = B^-_k \left(\widehat{\omega}_k \right) = 0,
\end{equation}
where
\begin{subequations}
\begin{equation}
B^+_k(f) = \int_0^1 e^{-\vert \kappa \vert y} f(y)\mathrm{d}y, \quad B^-_k(f) = \int_0^1 e^{- \vert \kappa \vert(1-y)} f(y) \mathrm{d}y = 0,
\end{equation}
for $ k \neq 0$, and
\begin{equation}
B^+_0(f) = \int_0^1 y f(y) \mathrm{d}y = 0, \quad \mathrm{and} \quad B^-_0(f) = \int_0^1 (1-y) f(y)\mathrm{d}y = 0.
\end{equation}
\end{subequations}
For numerical purposes, it is better to reformulate these stiff conditions by
taking advantage of the diffusion operator, i.e. by applying $B^+$ and $B^-$ to (\ref{eq:vorticity_transport}),
which leads to
\begin{equation}
\label{eq:integral_bc}
B^\pm_k(\partial_y^2 \widehat{\omega}_k) = \nu^{-1} B^\pm_k\left(\widehat{\nab(\uu \omega)}_k \right).
\end{equation}

The above analysis ensures that the system of equations (\ref{eq:vorticity_transport}), (\ref{eq:vorticity_ic}), (\ref{eq:psi_def}), (\ref{eq:psi_explicit}) and either one of (\ref{eq:integral_bc_old}) or (\ref{eq:integral_bc})
is equivalent to the original Navier--Stokes system for smooth strong solutions.

	\subsection{Prandtl--Euler model}

We now describe the alternative model for the flow derived by \cite{Prandtl1904}.
Although Prandtl and most later authors used the velocity variable to write down the equations,
we present here the equivalent vorticity formulation,
since we have found that it leads to a simpler understanding of the phenomena we are interested in.

The starting point is the following Ansatz for the vorticity field as $\Rey \to \infty$:
\begin{equation}\label{eq:prandtl_ansatz}
\omega(x,y) = \omega_E(x,y) + \nu^{-1/2} \omega_P(x,\nu^{-1/2}y) - \nu^{-1/2} \omega_P(x,\nu^{-1/2}(1-y)) + \omega_R(x,y),
\end{equation}
where $\omega_E(x,y)$ is a smooth function on $\Omega \times ]0,T[$, $\omega_P(x,y_P)$ is a smooth function on $C = \TT \times ]0,\infty[ \times ]0,T[$
which decays rapidly when $y_P \to \infty$.
The indices $_E, _P$ and $_R$ denote respectively the Euler, Prandtl and remainder terms, and $y_P = \nu^{-1/2} y$ is the Prandtl variable. 
%
Note that for simplicity, we have assumed that the flow is symmetric around the channel axis,
so that the two $\omega_P$ terms correspond to two symmetric BLs of opposite sign at $y=0$ and $y=1$.

By a classical multiple scales analysis, it can be formally shown that $\omega_E$ should satisfy the incompressible Euler equations in $\Omega$, 
and $\omega_P$ the Prandtl equations,
\begin{subequations}\label{eq:Prandtl}
\begin{align}
& \label{eq:prandtl_transport} \partial_t \omega_P + \nab ( \uu_P \omega_P) = \partial_{y_P}^2 \omega_P \\
& \omega_P(x,y_P,0) = 0 \\
& \psi_{P}(x,y_P,t) = \int_0^{y_P} \mathrm{d}y_P' \int_0^{y_P'} \mathrm{d}y_P'' \omega_P(x,y_P'',t)  \\
& \label{eq:Prandtl_bc_2} \partial_{y_P}\omega_P(x,0,t) = -\partial_x p_E(x,0,t),
\end{align}
\end{subequations}
where $p_E$ is the pressure field calculated from $\omega_E$.
It is instructive to rederive the classical Neumann condition (\ref{eq:Prandtl_bc_2}) as follows.
First, by replacing $\omega$ according to (\ref{eq:prandtl_ansatz}) in (\ref{eq:integral_bc_old}),
one obtains:
\begin{equation}\label{eq:prandtl_exact_bc}
\int_0^1 \widehat{\omega}_{Ek}(y) \exp(-\vert \kappa \vert y)\mathrm{d}y + \nu^{-1/2}\int_0^1 \widehat{\omega}_{Pk}(y) \exp(-\vert \kappa \vert y)\mathrm{d}y \, = \, 0 \, ,
\end{equation}
and by expressing the second integral with respect to $y_P$ and keeping the lowest order term in $\nu$:
\begin{equation}
\int_0^\infty \widehat{\omega}_{Pk}(y_P)\mathrm{d}y_P = -\int_0^1 \widehat{\omega}_{Ek}(y) \exp(-\vert \kappa \vert y)\mathrm{d}y
\end{equation}
Then by integrating (\ref{eq:prandtl_transport}) over $[0,\infty]$, 
one finds that the contribution of the nonlinear term vanishes, 
and is left with
\begin{equation}\label{eq:prandtl_integral_bc}
-\partial_t \int_0^1 \widehat{\omega}_{Ek}(y) \exp(-\vert \kappa \vert y)\mathrm{d}y = \int_0^{\infty} \partial_{y_P}^2 \widehat{\omega}_{Pk}(y) = - \partial_{y_P} \widehat{\omega}_{Pk}(0)
\end{equation}
where, from the considerations in the preceding paragraph, 
it appears that the left-hand side can be identified with the pressure gradient $\partial_x p_E(x,0,t)$.
Intuitively, the wall pressure gradient computed from the Euler solution creates vorticity at the boundary,
which then diffuses inwards and evolves nonlinearly due to the flow it generates in the BL.

Since the Prandtl equations do not include diffusion parallel to the wall,
nothing prevents in general that the vorticity gradient in the $x$ direction grows indefinitely,
hence the possibility of finite time singularity.
More precisely, the mechanism proposed by \cite{VanDommelen1990}
is that a fluid element is compressed to a point in the wall-parallel direction,
and extends to infinity in the wall-normal direction.
We shall denote by $t^*$ the time at which this first occurs,
and by $x^*$ the corresponding $x$ location.
From the scaling exponents computed by \cite{VanDommelen1990},
we can deduce that if the initial data are analytic, 
the spectrum of the solution will fill when approaching singularity 
with a characteristic cutoff parallel wavenumber scaling like
\begin{equation}\label{eq:wavenumber_blowup}
k_C \propto (t^*-t)^{-3/2}.
\end{equation}

	\subsection{Interactive boundary layer model}

As the singularity builds up in the Prandtl solution, 
the corresponding Navier--Stokes solution adopts a quite different behavior.
As first explained by \cite{Elliott1983}, the first divergence between the two solutions
occurs when the outer potential flow generated by the BL vorticity 
creates a pressure gradient perturbation of order 1 at the wall,
which in turn impacts the inward diffusion of vorticity.
This effect generically starts to take place when
\begin{equation}\label{eq:interactive_perturbation_time}
t^* - t = O \left(\Rey^{-2/11} \right).
\end{equation}

A rigorous asymptotic description of this new effect would require the modification of the vorticity ansatz (\ref{eq:prandtl_ansatz})
with new BLs, both in $x$ and in $y$, coming into play.
To avoid such complications, we follow \cite{Peridier1991a} 
and consider the finite Reynolds number description called interactive boundary layer (IBL) model,
which simply consists in modifying the Prandtl equations to include the new large scale interaction,
but without trying to rescale the solution {\it a priori}.
Ansatz (\ref{eq:prandtl_ansatz}) therefore remains valid, except that $\omega_P$ is replaced by $\omega_I$, the solution of the interactive equations
which we shall now derive.

Since we are working with the vorticity formulation, we are blind to potential flow perturbations,
but their effect manifests itself through the integral boundary condition (\ref{eq:integral_bc_old}) on $\omega$.
Starting again from (\ref{eq:prandtl_exact_bc}), but expanding  the exponential up to order $\Rey^{-1/2}$, yields
\begin{equation}
\int_0^1 \widehat{\omega}_{Ik}(y_P) \mathrm{d}y_P - \nu^{1/2} \vert \kappa \vert \int_0^\infty  y_P \widehat{\omega}_{Ik}(y_P) \mathrm{d}y_P
= - \int_0^1 \widehat{\omega}_{Ek}(y) \exp(-\vert \kappa \vert y)\mathrm{d}y
\end{equation}	
and, following the same procedure as above, leads to a perturbed boundary condition for $\omega_I$:
\begin{equation}\label{eq:interactive_bc}
\partial_{y_P} \widehat{\omega}_{Ik}(0) = -\iota \kappa \widehat{p}_{Ek}(0) - \nu^{1/2} \vert \kappa \vert \partial_t \widehat{\beta}_{Ik},
\end{equation}
where
\begin{equation}
{\beta}_{Ik}(x) = \int_0^\infty y_P {\omega}_{I}(x,y_P) \mathrm{d}y_P.
\end{equation}
On the other hand, by multiplying (\ref{eq:prandtl_transport}) by $y_P$ and integrating over $y_P$, we obtain an expression for $\partial_t \beta_I$,
which closes the problem.

As a side remark, let us note that the classical name ``interactive boundary layer'' for this model is misleading, 
since in fact no retroaction of the Prandtl layer onto the bulk Euler flow is taken into account.
An alternative name could be ``wet boundary layer'', 
which better encompasses the notion that the potential far flow affects the boundary layer equations only through a passive effect.

	\subsection{Orr--Sommerfeld model}\label{sec:orr-sommerfeld-model}

Several numerical studies suggest that a linear instability mechanism could play a role in the detachment process.
Since we are concerned with an unsteady problem,
the notion of linear instability should be understood here in an asymptotic sense, 
in terms of a rescaled time variable in which the evolution of the base flow can be neglected.
Moreover, since we are looking for an instability happening at high wavenumbers in the parallel direction,
we also neglect, for the time being, the parallel variation of the base flow,
or in other words we study the possible occurence of perturbations which have a large parallel wavenumber $k$ compared to 
the characteristic parallel wavenumber $L^{-1}$ of the flow prior to detachment.
The combination of both hypotheses constitutes the frozen flow approximation.
Its domain of validity could be properly evaluated only by resorting to a multiple time scale asymptotic analysis, which we have not yet achieved in this setting.

	\subsubsection{Formulation}

Under these two simplifying hypotheses, we are brought back to Rayleigh's classical shear flow stability problem, 
later generalized to viscous fluids by Orr and Sommerfeld. 
In the case of a boundary layer, several simplifications are possible
which allowed \cite{Tollmien1929} to obtain an elegant asymptotic description of 
the modes now known as Tollmien--Schlichting waves, 
and of the corresponding stability region in the $(\Rey,k)$ plane, 
which was later confirmed experimentally by \cite{Schubauer1948}.
For a more recent review on the subject, see \cite{Reed1996}.
Although most of the material presented here is classical (see \cite{Lin1967}),
previous studies have mostly emphasized the computation of the critical Reynolds number,
so that it is instructive to rederive the main results directly in the $\Rey \to \infty$ limit, which concerns us here.	

For small perturbations $\delta\psi(x,y_P,t_2) = \phi(y_P)e^{i(\kappa x-\alpha t_2)}$ to the stream function,
the profile function $\phi$ satisfies the Orr--Sommerfeld equation
\begin{equation}\label{eq:orr-sommerfeld}
(u_P-c)(\phi'' -  \nu\kappa^2 \phi) - u_P'' \phi = \frac{1}{i \kappa} (\phi'''' - \nu\kappa^2 \phi'' + \nu^2\kappa^4\phi),
\end{equation}
where $c = \alpha/\kappa$ is the phase velocity, and primes denote derivatives with respect to $y_P$.
Note that $c$ is in general a complex number,
and unstable perturbations are those 
for which $c$ has a strictly positive imaginary part.
Now assuming that
\begin{equation}\label{eq:k_ordering}
L^{-1} \ll k \ll L^{-1}\Rey^{1/2},
\end{equation}
(\ref{eq:orr-sommerfeld}) simplifies to
\begin{equation}
\label{eq:tollmien2}
(u_P-c)\phi'' - u_P''\phi = \frac{1}{i\kappa} \phi''''.
\end{equation}
The no-slip boundary condition translates to $\phi(0) = \phi'(0) = 0$.
Assuming from now on that $k>0$ without loss of generality, 
the boundary condition for $y_P \to +\infty$ can be obtained by matching $\phi$ 
with a harmonic outer solution of the form $\exp(-\kappa y)$ using the hypothesis that vorticity vanishes outside the BL, 
which means that 
\begin{equation}
\label{eq:cond_inf}
\phi(y_P) \underset{y_P \to \infty}{=} A(1-\kappa \nu^{1/2}  y_P) + o(\kappa \nu^{1/2} y_P).
\end{equation}
Note that it is essential to keep the first order term in this expression in order to find unstable modes.
Following \cite{Tollmien1929}, we now deal with the singular perturbation problem (\ref{eq:tollmien2})
by first considering inviscid solutions, and then adding a boundary layer.

	\subsubsection{Inviscid mode}

Neglecting the viscous contribution in (\ref{eq:tollmien2}), we obtain
\begin{equation}\label{eq:tollmien1}
(u_P-c)\phi'' - u_P'' \phi = 0
\end{equation}
which admits the obvious regular solution
\begin{equation}\label{eq:def_phi_1}
\phi_{1,c} = u_P-c,
\end{equation}
but is singular in any point where $u_P = c$.
To construct another independent solution $\phi_{2,c}$, we now assume that $c$ does not lie directly on the real axis, 
and we make the change of unknown
$
\phi = (u_P-c)f,
$
leading to 
\begin{equation}\label{eq:phi_2_eq}
(u_P-c)f'' + 2u_P' f' = 0
\end{equation}
and thus
\begin{equation}
\phi_{2,c}(y_P) = (u_P-c)\left(\int_\infty^{y_P}  \left(\left(\frac{u_\infty-c}{u_P-c}\right)^2 - 1\right) + y_P\right) \, ,
\end{equation}
where $u_\infty$ is the velocity outside the boundary layer.
By combining $\phi_{1,c}$ and $\phi_{2,c}$,
a solution $\phi_{out}$ satisfying the condition (\ref{eq:cond_inf}) at $+\infty$ is readily obtained:
\begin{equation}\label{eq:def_phi_out}
\phi_{out} = \phi_{1,c} - \kappa\nu^{1/2} \phi_{2,c} \, .
\end{equation}

	\subsubsection{Viscous correction} \label{sec:viscous_correction}

We now look for a viscous sublayer correction $\phi_{in}$ which is necessary since $\phi_{out}$ does not in general satisfy the no-slip boundary condition at $y_P = 0$.
For small $y_P$, (\ref{eq:tollmien2}) reduces to 
\begin{equation}\label{eq:tollmien4}
u_P'(z_0(c))(y_P-z_0(c))\phi'' - u_P''(0) \phi(0) = \frac{1}{i\kappa} \phi'''',
\end{equation}
where we have defined $z_0(c)$ to be the solution with the smallest real part to the equation $u(z)=c$ (see Appendix \ref{sec:I_c_appendix}).
An inner variable can then be defined in the viscous sublayer by  $\eta = \varepsilon\left(y_P-z(c)\right)\left\vert \kappa  u_P'(z(c))  \right\vert5^{1/3}$,
where $\varepsilon = \mathrm{sign}(u_P'(z(c)))$,
leading to with $\kappa \gg 1$ to
\begin{equation}
\label{eq:tollmien3}
\eta \widetilde{\phi}'' = -i \widetilde{\phi}'''',
\end{equation}
We are interested in a solution of this equation which remains bounded and whose derivative tends to zero when $y_P \to \infty$.
When $\varepsilon > 0$, this limit is equivalent to $\eta \to \infty$, and a solution to the problem was given by \cite{Tollmien1929} in terms of Hankel functions:
\begin{equation}
\widetilde{\phi}_{in}(\eta) = \int_\eta^\infty \mathrm{d}\eta' \int_{\eta'}^\infty \mathrm{d}\eta'' {\eta''}^{1/2} H^{(1)}_{1/3}\left(\frac{2}{3}(i\eta'')^{3/2}\right)
\end{equation}
Note that, as long as it is expressed in terms of the $\eta$ variable,
this solution is universal.
Expressed as a function of $y_P$, it reads
\begin{equation}\label{eq:phi_in}
\phi_{in}(y_P) = \widetilde{\phi}_{in}\left( \left(y_P-\frac{c}{u_P'(0)}\right)\left\vert\kappa u_P'(0)\right\vert^{1/3} \right)
\end{equation}
In the case $\varepsilon < 0 $, $y_P \to \infty$ corresponds to $\eta \to -\infty$,
and the solution $\widetilde{\phi}_{in}(\eta)$ should be adapted accordingly. 

	\subsubsection{Construction of unstable modes}

Now in the asymptotic regime, any admissible solution $\phi$ of (\ref{eq:tollmien2}) can approximately be expressed as
\begin{equation}
\phi = A \phi_{out} + B \phi_{in},
\end{equation}
so that the boundary conditions $\phi(0) = \phi'(0) = 0$ translate to the linear system
\begin{equation}
\begin{cases} 
A \phi_{out}(0) + B \phi_{in}(0) = 0 \\
A \phi'_{out}(0) + B \phi'_{in}(0) = 0
\end{cases}.
\end{equation}
In order for a nontrivial solution to exist,
this system should be degenerated, i.e.
\begin{equation}
\label{eq:degeneracy_condition}
\frac{\phi_{in}(0)}{\phi_{in}'(0)} = \frac{\phi_{out}(0)}{\phi_{out}'(0)}.
\end{equation}

On the one hand, denoting 
\begin{equation}\label{eq:def_eta_w}
\eta_w = - c\kappa^{1/3}\left\vert u_P'(0) \right\vert^{-2/3}
\end{equation}
the position of the wall in terms of the $\eta$ variable, it is shown following \cite{Tollmien1929} that
%
\begin{equation}\label{eq:tietjens}
\frac{\phi_{in}(0)}{\phi_{in}'(0)} = -\frac{c}{u_P'(0)} F(-\eta_w) \mathrm{\quad if \,} u_P'(0) > 0 ,
\end{equation}
where $F$ is a one-parameter complex function known as the Tietjens function $F$.
Although there is no closed analytical formula for $F$, it is easily approximated from (\ref{eq:phi_in}) using quadrature formulas.
\begin{figure}
\begin{center}
\includegraphics[width=0.6\columnwidth]{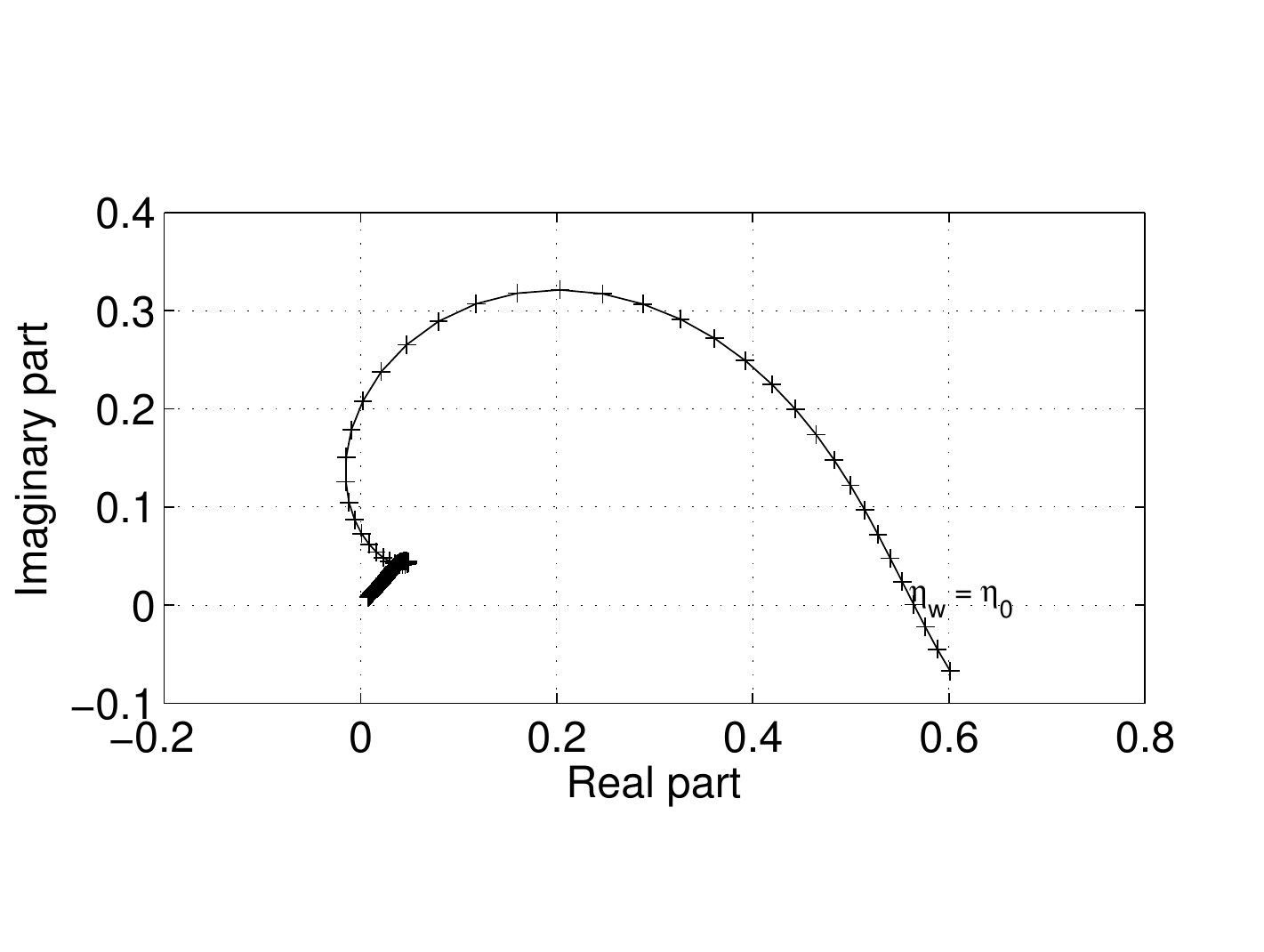}
\end{center}
\caption{
\label{eq:tollmien_function}
Imaginary vs. real part of the Tietjens function $F(\eta)$ for $\eta \in [2,20]$.
The sampling points (every $0.1$) are indicated by crosses.
}
\end{figure}
A graphical representation is given in Fig.~\ref{eq:tollmien_function}.

On the other hand, 
denoting for convenience of notation
\begin{equation}\label{eq:I_c_def}
I_c = -c \int_0^\infty \left(\frac{1}{(u_P(y)-c)^2} - \frac{1}{(u_\infty-c)^2}\right) \mathrm{d}y
\end{equation}
and using (\ref{eq:def_phi_out}), it is shown that
\begin{equation}
\label{eq:def_E}
\frac{\phi_{out}(0)}{\phi_{out}'(0)} = -\frac{c}{u_P'(0)}E(\kappa\nu^{1/2},c),
\end{equation}
where
\begin{equation}
\label{eq:outer_ratio}
E(\kappa\nu^{1/2},c) = 1 - \frac{\kappa\nu^{1/2} (u_\infty-c)^2}{c u_P'(0) \left(1 - \frac{\kappa\nu^{1/2}(u_\infty-c)^2}{c}(I_c-\frac{1}{u'(0)}) \right)}.
\end{equation}

Injecting (\ref{eq:tietjens}) and (\ref{eq:def_E}) back into the degeneracy condition (\ref{eq:degeneracy_condition}),
we obtain the dispersion relation 
\begin{equation}
F(-\eta_w) = E(\kappa\nu^{1/2},c).
\end{equation}
The profile is linearly unstable if and only if this equation has solutions such that $Im(c) > 0$.
In the range of $\kappa$ we are considering, there can be no solutions if $c$ is of order $u_\infty$ or larger, 
because then $\eta_w \to -\infty$, and therefore $F(-\eta_w) \to 0$, whereas $E$ remains of order $1$.
In the following, we therefore look for solutions under the restriction that $c << u_\infty$.

The asymptotic behavior of $I_c$ for small $c$ is dominated by what happens near solutions of $u(y_P)=0$.
As established in Appendix \ref{sec:I_c_appendix}, if $u_P$ and $u_P'$ do not have any zeros in common, then
\begin{equation}\label{eq:I_c_estimate}
I_c \underset{c\to 0}{=} \frac{1}{u_P'(0)} - c \frac{u_P''(0)}{u_P'(0)^3} \ln\left(-\frac{c}{u_P'(0)} \right) -c i \pi \Delta_u  + c K_R + o(c)
\end{equation}
where $K_R$ is a real constant, $\ln$ is the principal branch of the complex logarithm, and
\begin{equation}
\Delta_u = \sum_{\{y_P > 0 \mid u_P(y_P)=0\}} \frac{u_P''(y_P)}{u_P'(y_P)^3}.
\end{equation}

The behavior of this asymptotic expression when $c$ approaches the real axis is tricky and should be considered with care.
If $\Re(c/u'_P(0)) < 0$, the argument of the complex logarithm lies in the right hand side of the complex plane,
and the limit $\Im(c) \to 0$ is well behaved.
But if $\Re(c/u'_P(0)) > 0$, the limit does not exist strictly speaking, and $I_c$ should be considered as multi-valued,
which is well captured by replacing $\Delta_u$ by
\begin{equation}
\Delta^\pm_u = \begin{cases} \Delta_u \mathrm{\,\, if\,\,} \Re(c/u'_P(0)) < 0 \\ 
\pm \frac{u_P''(0)}{u_P'(0)^3} + \Delta_u \mathrm{\,\, if\,\,} \Re(c/u'_P(0)) > 0
\end{cases}.
\end{equation}

By injecting our estimate for $I_c$ in (\ref{eq:outer_ratio}), and assuming for simplicity that $\kappa \nu^{1/2}\log\vert c\vert \ll 1$,
we obtain the following estimates for the real and imaginary parts of $E$ (respectively)
when $c$ is close to the real axis:
\begin{subequations}
\label{eq:estimates_E}
\begin{align}
\label{eq:estimate_E_real}
\Re(E) & \sim 1 - \frac{\kappa\nu^{1/2} u_\infty^2}{c u_P'(0)}\\
\label{eq:estimate_E_imag}
\Im(E) & \sim \frac{\kappa^2 \nu u_\infty^4 \Delta^\pm_u \pi}{c u_P'(0)},
\end{align}
\end{subequations}
where we have assumed that $\Delta^\pm_u \neq 0$
(this analysis therefore does not apply directly to the Blasius boundary layer).
Since the imaginary part of $E$ is negligible compared to its real part,
(\ref{eq:degeneracy_condition}) can be satisfied only in the neighborhood
of points where $F$ is purely real.
This happens for $\eta_w \to -\infty$, as well as for a certain finite value $\eta_0 \simeq -2.3$
where $F(-\eta_0) \simeq 0.56$ (see Fig.~\ref{eq:tollmien_function}).

In the case $u'_P(0) < 0$, the latter leads, with (\ref{eq:estimate_E_real}), to a single critical wavenumber
\begin{equation}
\label{eq:critical_k_1}
k_1 \simeq 0.161 \, \vert u_P'(0) \vert ^{5/4} u_\infty^{-3/2} \nu^{-3/8},
\end{equation}
beyond which the modes are unstable.
In the case $u'_P(0) > 0$, the multi-valuedness of $I_c$ implies that $k_1$ splits into two critical wavenumbers
with very close values, and therefore, a negligible instability region.

The critical point near $\eta_w \to -\infty$, 
corresponds to $F(-\eta_w) \to 0$,
so from (\ref{eq:estimate_E_real})
\begin{equation}
\frac{\kappa\nu^{1/2} u_\infty^2}{c u_P'(0)} \simeq 1.
\end{equation}
To obtain another equation relating $c$ and $\kappa$, we use the estimate
$
F(-\eta_w) \underset{\eta \to -\infty}{\sim} -e^{i\pi/4} \vert \eta \vert^{-3/2},
$
which, combined with (\ref{eq:estimate_E_imag}), leads to
\begin{equation}
\frac{\kappa^2 \nu u_\infty^4 \Delta^\pm_u \pi}{c u_P'(0)} \simeq \frac{\sqrt{2}}{2} \vert \eta_w \vert^{-3/2}.
\end{equation}
This equation has a simple root iff $\Delta_u > 0$ and $u'_P(0) < 0$, in which case we obtain a second critical wavenumber
\begin{equation}
\label{eq:critical_k_2}
k_2 = 0.0968 \, \vert u_P'(0) \vert^{1/2} \vert \Delta_u \vert^{-1/3} \vert u_\infty \vert^{-5/3} \nu^{-5/12}
\end{equation}
In other cases, if there exists an unstable region for $k > k_1$,
its upper bound cannot be found under our current restriction $k \ll \nu^{-1/2}$,
which implies that it extends at least up to wavenumbers scaling like $\nu^{-1/2}$,
corresponding to what is usually called the Rayleigh instability.
This observation should be kept in mind as it is one of the key elements of the detachment scenario we will propose further down.

Another important quantity we need to estimate is the growth rate of the unstable modes.
From (\ref{eq:outer_ratio}), we see that since $c$ remains small in absolute value,
the growth rate $\alpha$ of the instability satisfies
\begin{equation}\label{eq:growth_rate}
\alpha \sim \frac{u_\infty^2 \Im (F(-\eta_w)) }{\vert u'_P(0) \vert}\nu^{1/2}\kappa^2
\end{equation}

To sum up, the generic instability expected to play a role in such boundary layer flows in the inviscid limit
manifest itself by the growth of wave packets in the vicinity of the boundary confined in physical space
to regions where $u'_P(0) < 0$ (recirculation bubbles),
and whose parallel wavenumber support extends from $O(\Rey^{3/8})$ to at least $O(\Rey^{1/2})$.

	\subsubsection{The case $u_P'(0) = 0$}

To be complete our analysis should also take into account the case $u_P'(0) = 0$,
 investigated in detail by \cite{Hughes1965}.
Going back to the general expression (\ref{eq:def_phi_out}) for the outer solution,
we obtain in the case $u_P'(0) = 0$ that
\begin{equation}
\label{eq:outer_ratio2}
E(\kappa\nu^{1/2},c) = \frac{\phi_{out}(0)}{\phi_{out}'(0)} = -\frac{c^2}{ \kappa \nu^{1/2} (u_\infty-c)^2 } - c I_c
\end{equation}
or, with (\ref{eq:I_c_up_0}), and when $c$ is close to the real axis:
\begin{subequations}
\begin{align}
\Re(E) & \simeq -\frac{c^2}{ \kappa \nu^{1/2} u_\infty^2 }\\
\label{eq:E_imag}
\Im(E) & \simeq -\frac{\pi}{4} \left(\frac{2c}{u_P''(0)}\right)^{1/2}
\end{align}
\end{subequations}

Concerning the inner solution, (\ref{eq:tietjens}) is replaced by
\begin{equation}
\frac{\phi_{in}(0)}{\phi'_{in}(0)} = -z(c) F(-\eta_w),
\end{equation}
which, combined with (\ref{eq:E_imag}), yields
\begin{equation}
\Im (F(-\eta_w)) = -\frac{\pi}{4},
\end{equation}
or equivalently, according to \cite{Hughes1965},
$$
\eta_w \simeq -0.488, \quad \Re(F(-\eta_w)) \simeq 1.580,
$$
and therefore
\begin{align}
& \frac{c^{3/2}}{ \kappa \nu^{1/2} u_\infty^2 } = \sqrt{\frac{2}{u_P''(0)}} 1.580, \\
& \frac{2c}{u_P''(0)} \kappa^{1/3} (2cu''(0))^{1/3} = 0.488.
\end{align}
%
which finally gives us the critical wavenumber
$$
k_1 = 0.0279 \, u_P''(0)^{10/11} u_\infty^{16/11} \nu^{-4/11} 
$$

The growth rate, obtained following the same reasoning as above 
with (\ref{eq:outer_ratio}) replaced by (\ref{eq:outer_ratio2}), reads
\begin{equation}\label{eq:growth_rate2}
\alpha \sim u_\infty \Im (F(-\eta_w)) \nu^{1/4}\kappa^{3/2}
\end{equation}

	\subsubsection{Physical interpretation}

In this section we formulate some conjectures relevant to the physical interpretation of the above model. 
We have shown that, subject to the validity of the frozen flow approximation, 
all BL flows containing recirculation bubbles are subject to Tollmien--Schlichting--Rayleigh instabilities
for wavenumbers $k \in [k_1,k_2]$, where $k_1$ and $k_2$ both diverge to $\infty$ when $\Rey \to \infty$.

Therefore a plausible scenario for detachment may begin as follows.
Suppose that initially the flow is very smooth, for example, that it has analytic regularity,
i.e. its Fourier coefficients decay exponentially with $k$,
and that a recirculation bubble appears due to the Prandtl BL nonlinear dynamics.
Even though the range $[k_1,k_2]$ is already subject to an instability,
for sufficiently large Reynolds number the initial excitation of such high wavenumber modes is so small
that they do not have time to grow and the Prandtl solution is a good approximation.

But if and when a Prandtl singularity builds up, it starts feeding non negligible excitations
into the interval $[k_1,k_2]$,
In the competition between the oncoming singularity and the growth of unstable modes,
it is interesting to determine which modes first reach a finite amplitude, and when this occurs.

Now if we replace $\kappa$ by the characteristic excitation (\ref{eq:wavenumber_blowup}) generated by the Prandtl dynamics
some time $t^*-t$ before the singularity, we obtain
\begin{equation}\label{eq:effective_growth_rate}
\alpha \propto \nu^{1/2} (t^*-t)^{-3}.
\end{equation}
With this growth rate, the first perturbations to reach order 1 occur at a time
\begin{equation}\label{eq:instability_perturbation_time}
t^*-t = O(\Rey^{-1/4}).
\end{equation}
By comparing this result with (\ref{eq:interactive_perturbation_time}), we note that this occurs later than the perturbations due to large scale interactions,
as described by the IBL model.
Therefore, the BL profile resulting of an IBL computation,
not the Prandtl profile, should be used as base profile when performing the stability analysis.
This confirms the analysis of \cite{Gargano2014}, who pointed out that what they call a \textit{large scale interaction}
always preceeds the approach to detachment.

In the region with reversed flow near the wall,
the unstable wavenumber range scales likes $O(\Rey^{1/2})$.
Assuming that all the modes grow simultaneously and reach order 1,
this means that the support of $\widehat{\omega}$ extends to $k \propto O(\Rey^{1/2})$,
while the amplitude of the modes continues to scale like $O(\Rey^{1/2})$.
Due to the properties of the inverse Fourier transform,
these scalings immediately imply that the profile of $\omega$ very near the wall 
has a kind of wave packet shape with amplitude scaling like $O(\Rey^1)$ indeed.

During the linear phase, the characteristic wall-normal extent of such modes 
is controlled by the considerations of Sec.~\ref{sec:viscous_correction},
i.e. $\kappa^{-1/3} \Rey^{-1/2} \sim \Rey^{-2/3}$.
But once the unstable modes have reached order 1 
and the amplitude of $\omega$ scales like $O(\Rey^1)$
(due to the superposition of all modes as noted above),
nonlinear vorticity advection effects imply that 
the characteristic scale becomes $O(\Rey^{-1})$,
which gives us a possible physical explanation for the Kato layer.

\section{Solvers}

	\subsection{Setup}

To trigger an unsteady separation process,
we have chosen an initial configuration inspired by the dipole of \cite{Orlandi1990},
later modified by \cite{Clercx2002}.
However, this dipole has the drawback of generating a secondary, weaker dipole propagating in the opposite direction which is computed at a waste.
For efficiency reasons, we have thus preferred a quadrupole configuration, which is symmetric both around the channel axis
and around the midplane, thus sparing $3/4$ of the domain size for a given $\Rey$.
It is defined in terms of its stream function as follows:
\begin{subequations}
\begin{align}
\label{eq:initial_psi}
\psi_i(x,y) & = A x y \exp\left(-\frac{(x-x_0)^2 + (y-y_0)^2}{2s^2}\right),
\end{align}
\end{subequations}
%
where $A = 0.625847306637464$ determines amplitude of the vortices,
$s =  6.3661977236758$ their size and $(x_0,y_0) = (0.5,0.5)$ their initial location.
Note that the boundary conditions are satisfied only approximately by this velocity field, 
but in fact
$$
v_{i}(x,y=0) \approx 10^{-15}. 
$$
which is anyway of the same order as the round-off error in double precision arithmetics.

\begin{figure}
\begin{center}
\includegraphics[width=0.6\columnwidth]{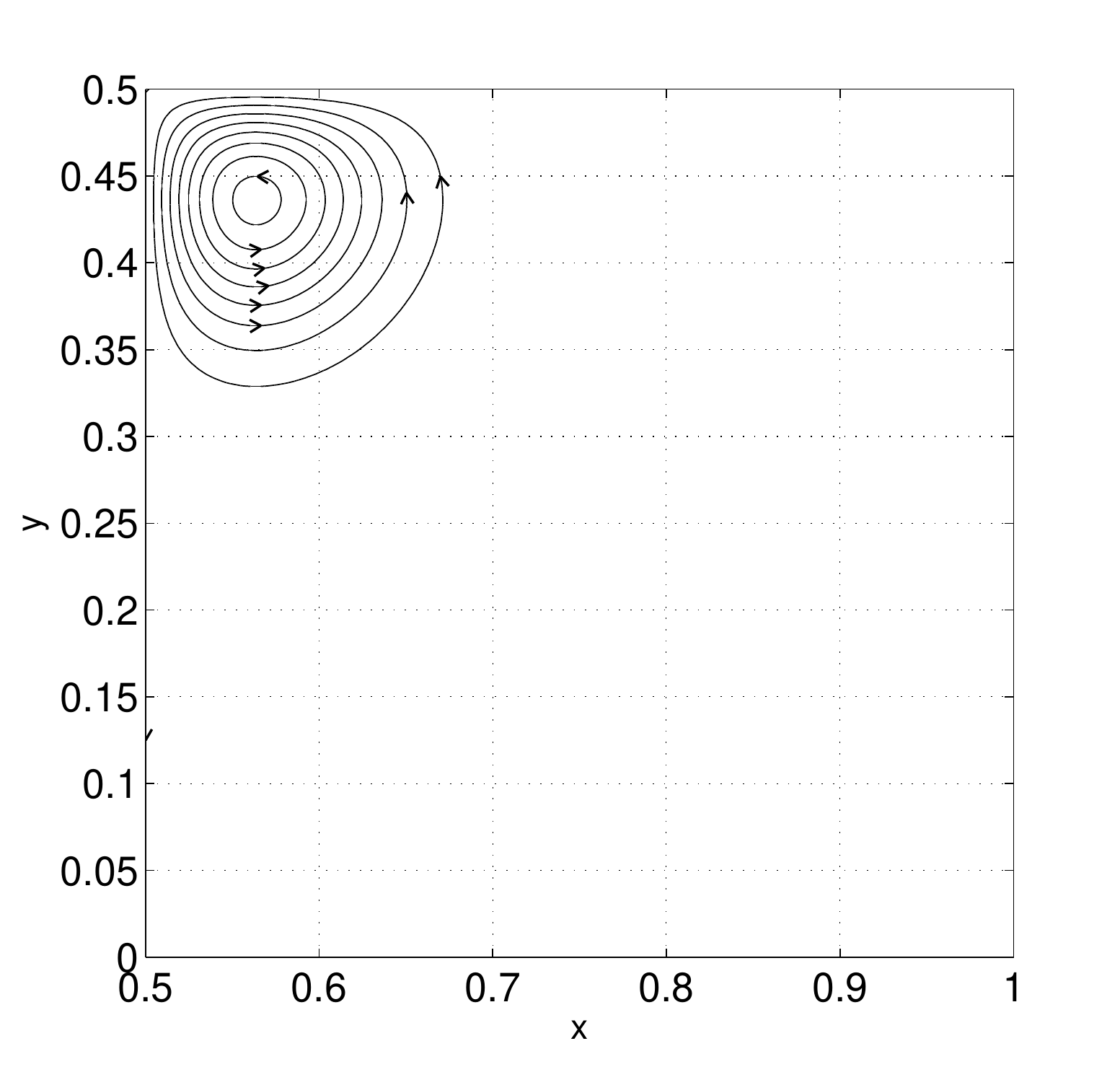}
\end{center}
\caption{
\label{fig:initial_data}
Streamlines of initial velocity field in the subdomain $K = [1/2,1] \times [0,1/2]$.
}
\end{figure}

Due to the symmetry of this initial condition, the analysis can be restricted without loss of information to the subdomain $K = [0,1/2] \times [0,1/2]$.
The streamlines of $\uu_i$ in $K$ are shown in Fig.~\ref{fig:initial_data}.
The definitions and initial values of several integral quantities which we will be important in our study are given in Table~\ref{tab:integral_quantities}.

\begin{table}
\begin{center}
\begin{tabular}{ccccc}
quantity & enstrophy & maximum vorticity & energy & maximum velocity \\
definition & $\displaystyle{\frac{1}{2} \int_\Omega \omega^2}$ & $\displaystyle{\max_\Omega \omega}$ 
& $\displaystyle{\frac{1}{2} \int_\Omega \vert \mathbf{u} \vert^2}$ & $\displaystyle{\max_\Omega \vert \mathbf{u} \vert}$ \\
initial value & $7.48 \cdot 10^{-3}$ & $1.00$ & $9.05 \cdot 10^{-6}$ & $2.42 \cdot 10^{-2} $ \\
\end{tabular}
\end{center}
\caption{
\label{tab:integral_quantities}
Definitions and initial values of several integral quantities of interest.}
\end{table}

In this work we shall analyze the flows obtained by solving the Navier--Stokes equations numerically up to $T=57.05$
for $\nu$ decreasing from $4\cdot 10^{-7}$ to $5 \cdot 10^{-8}$ by factors of $\sqrt{2}$ (i.e., $7$ different values in total).
Reynolds numbers corresponding to these values of $\nu$ are defined according to (\ref{eq:def_reynolds}),
where $U \simeq 2.42 \cdot 10^{-2} $ is the initial maximum velocity, and $L= 2s \simeq 1.27 \cdot 10^{-1}$ is a measure of the size of the quadrupole.
Both $\nu$ and $\Rey$ are provided up to 3 significant digits in Table~\ref{tab:parameters}.
To facilitate comparison with previous results concerning dipole-wall collisions,
we have also included the Reynolds number $\Rey_{rms}$ computed from the RMS velocity $U_{rms} = 4.25 \cdot 10^{-3}$ and channel width instead,
which is of the same order of magnitude.

%
%
\begin{table}
\begin{center}
\begin{tabular}{cccccccccc}
 case 					& I 		& II 		& III 		& IV 		& V 	 & VI 		& VII 	 & VIII 	& IX 	 \\
 $10^8 \nu$ 			& $40.0$	& $ 28.3 $ 	& $ 20.0 $ 	& $14.1$& $10.0$ & $7.07$ & $ 5 $   & $3.54$	& $ 2.5 $	\\
 $10^{-2}\Rey_{rms}$ 	& $106$  	& $150$ 	& $213$ 	& $301$ & $425$  & $601$ & $851$ 	& $1200$	& $1700$ 	\\
 $10^{-2} \Rey$ 		& $76.8$  	& $109$ 	& $154$ 	& $217$ & $308$ & $435$   & $615$  	& $870$ 		& $1230$		\\
 $ 10^3 \Delta_{y_E,1}$ & $1.90$ 	& $1.91$ 	& $1.92$ 	& $1.92$& $1.93$ & $1.93$ & $1.93$ 	& $1.94$ 	& $1.94$ 	\\
 $ 10^5 \Delta_{y_P,1}$ & $15.4$ 	& $13.0$ 	& $11.0$ 	& $9.25$& $7.80$ & $6.57$ & $5.54$ 	& $4.66$ 	& $3.93$ 	\\
 $10^4 \Delta_{x,1}$ 	& $9.77$ 	& $9.77$	& $9.77$	& $9.77$& $9.77$ & $4.88$ & $4.88$	& $4.88$	& $4.88$ 	\\
 $ 10^3 \Delta_{y_E,2}$ & $1.76$ 	& $1.87$ 	& $1.95$ 	& $2.02$& $2.07$ & $2.12$ & $2.16$ 	& $2.19$ 	& $2.22$ 	\\
 $ 10^5 \Delta_{y_P,2}$ & $14.4$ 	& $12.7$ 	& $11.2$ 	& $9.71$& $8.39$ & $7.22$ & $6.18$ 	& $5.28$ 	& $4.50$ 	\\
 $ 10^5 \Delta_{y_K,2}$ & $11.4$ 	& $8.46$ 	& $6.24$ 	& $4.56$& $3.32$ & $2.40$ & $1.73$ 	& $1.24$ 	& $0.889$ 	\\ 
 $10^4 \Delta_{x,2}$ 	& $9.77$ 	& $9.77$	& $4.88$	& $2.44$& $1.22$ & $0.610$ & $0.305$	& $0.305$	& $0.153$
\end{tabular}
\end{center}
\caption{\label{tab:parameters}
Parameters of numerical experiments.
All figures in this table are given up to 3 significant digits.
}
\end{table}

\subsection{Navier--Stokes solver}

To solve the initial value problem for the Navier--Stokes equations, 
derivatives in the periodic $x$ direction are computed with spectral resolution from their sine and cosine series expansions.
Since the $y_P$ direction is not periodic, derivatives in the $y_P$ direction have to be treated differently.
The Chebychev scheme is accurate but very costly, and also imposes Gauss collocation points which are not optimal for our problem.
We have therefore prefered to turn to fifth order compact finite differences \citep{Lele1992,Gamet1999}.
Denoting by $(f_i)$ approximate values of a function $f$ on the uniform grid defined by $y_{1,i} = \frac{i}{N_{y_P}-1} (0 \leq i < N_{y_P})$,
and $(f_i')$, $(f_i'')$ approximations of its first and second derivatives at the same locations,
we impose fifth-order accuracy by requesting that 
\begin{subequations}
\begin{align}\label{eq:compact_fd}
\alpha_i f_{i-1}' + f_i' + \beta_i f_{i+1}' &= A_i f_{i-1} + B_i f_i + C_i f_{i+1} \\
\gamma_i f_{i-1}'' + f_i'' + \delta_i f_{i+1}'' &= D_i f_{i-1} + E_i f_i + F_i f_{i+1}
\end{align}
\end{subequations}
where the coefficients are calculated by matching the Taylor expansions of both sides up to fifth order.

Note however that these expressions are only valid for $1 \leq i < N_y-1$, 
so that two additional equations are needed to determine $(f_i')$ and $(f_i'')$  uniquely.
For the computation of $\partial_y(v \omega)$, 
they are obtained by noting that the derivative vanishes at $y=0$ and $y=1$, 
which is a direct consequence of the boundary conditions on $u$ and $v$ and of incompressibility.

For the viscous term $\partial_y^2 \omega$,
they should follow from the boundary conditions (\ref{eq:integral_bc}).
To derive them, the integrals are first discretized by a fifth-order local quadrature formula.
To preserve accuracy, $\omega$ is expanded locally into its Taylor polynomial form, 
and the contribution of the $k$-depdendent exponential factor is included using 
numerical algorithms for gamma functions from the \textsc{Boost.Math} library.
In order to  solve the two resulting square linear systems efficiently,
a parallel shared memory direct solver based on sparse LU factorization with pivoting is used,
as implemented in the \textsc{SuperLU} library \citep{Li1999,Demmel1999},
and the \textsc{PetscC} library \citep{Balay2013} is used for matrix arithmetics.
Note that due to the dependency of the integral constraint on $k$, 
the number of LU factorizations is multiplied by $K$.
The cost of these factorizations is considerable, and they are tractable only under the condition that $N_y$
is not too large.

To cope with the huge scale disparity between the bulk of the channel and the BL,
we therefore have to use non-uniform grids in the $y$ direction.
During the first phase of the flow evolution,
the BL is expected to follow Prandtl's scaling.
The total number of grid points is fixed to $N_{y,1} = 385$.
The grid spacing is set to a certain value $\Delta_{y_P,1}$ between 0 and 
$$
L_1 = 166 \, \Rey^{-1/2} L,
$$
which corresponds to the BL thickness as can be estimated from the Prandtl calculations,
and the remaining points are uniformly spread up to $y=1/2$, with spacing $\Delta_{y_E,1}$.

\begin{figure}
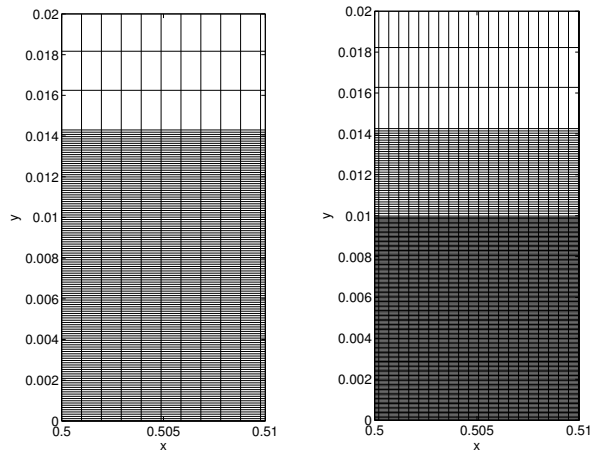

\begin{center}
\includegraphics[width=0.3\columnwidth]{{{FIGURES/grid1}}}
\includegraphics[width=0.3\columnwidth]{{{FIGURES/grid2}}} \\
\end{center}
\caption{
\label{fig:grids}
Schematics of discretization grids used by the Navier--Stokes solver for $t \leq 54$ (left) and $t > 54$ (right).
The thickness of the various regions is computed from the Reynolds number as summarized in Table.~\ref{tab:parameters}.
}
\end{figure}

At later times, the Prandtl scaling is expected to break down, and therefore a change of grid is required.
For convenience we always perform it at $t=54$ independently of $\Rey$.
The new grid has $N_{y,2} = 449$ points in the $y$ direction.
The grid spacing is set to $\Delta_{y_K,2}$ between 0 and 
$$
L_2 = 1210 \, \Rey^{-1} L,
$$
then to $\Delta_{y_P,2}$ between $L_2$ and $L_1$, and the remaining points are spread uniformly up to $y=1/2$.
The values of all deltas are given in Table~\ref{tab:parameters},
and graphical representations of the two grids are provided in Fig.~\ref{fig:grids}.

\subsection{Prandtl--Euler solver}

Following previous work \citep[see e.g.][]{Nguyenvanyen2009}, the Euler equation is approximated
by the Navier--Stokes equations with hyperdissipation,
\textit{i.e.}, the dissipation term $\nu \Delta \omega$ is replaced by $-\nu_2 (-\Delta)^2 \omega$.
This approximation is second order in space \citep{Kato1972}, which is sufficient for our purpose here.
The boundary conditions are enforced using the classical mirror image technique.
Since the vorticity field is antisymmetric with respect to $y=1/2$,
we just need to replace the boundary conditions at $y=0$ and $y=1$ by periodic boundary conditions
to effectively impose an exact non-penetration condition.
The Navier--Stokes equations are then solved on $\TT^2$, taking advantage of the symmetry of the solution, using a fully dealiased sine-cosine pseudo-spectral method corresponding to $N=4096$ grid points on the whole domain.
A low-storage third order Runge--Kutta scheme is employed for time discretization, the time step being adjusted dynamically to satisfy the CFL condition.
The hyperviscosity parameter $\nu_2$ was set to $2.146 \cdot 10^{-13}$, which was found to sufficiently regularize the solution.

To solve the initial value problem for the Prandtl equations (\ref{eq:Prandtl}), the spatial domain is first restricted to a finite size $L_{y_P}$ in the $y_P$ direction,
where $L_{y_P}$ should be chosen sufficiently large so that the solution does not depend on its value on the time interval considered.
The results presented below were obtained with $L_{y_P} = 64$.
Spatial discretization is then achieved as for the Navier--Stokes solver, except that the grid in $y$ is regular.

When computing the advection term $v_{P} \partial_{y_P} \omega_P$, 
the equations at the edges are obtained by shifting the stencils so that they remain inside the computational grid
(no additional condition is included).
For the dissipation term $\partial_{y_P}^2 \omega_P$, the integral constraint (\ref{eq:prandtl_integral_bc}) is rewritten
as follows:
\begin{equation}
\int_0^\infty \partial_{y_P}^2 \omega_P(x,y_P,t) \mathrm{d}y_P = \partial_x p_E(x,0,t),
\end{equation}
which is enforced as for the Navier--Stokes solver.
Finally, the system is closed by imposing that $\omega_P(x,L_{y_P},t) = 0$,
which is consistent with the fact that the exact solution decays rapidly in $y_P$.

\subsection{Interactive boundary layer solver}

The interactive solver is similar to the Prandtl solver, 
the only difference being that the pressure correction given by (\ref{eq:interactive_bc}) 
is included at each evaluation of the right hand side.
These modified boundary conditions unfortunately modify the stability region of
the time discretization scheme, making it much smaller.
We have heuristically derived the constraint 
$$
\Delta t < C \frac{\Delta_x^2}{\sqrt{\nu}},
$$
where $C = 1.5$.
For efficiency, we use the non-interactive Prandtl solver up to $t=50$,
and only then do we switch on the interactive term.

\subsection{Orr--Sommerfeld solver}

To compare the Navier--Stokes solution with the predictions of our linear instability model
beyond asymptotics, we have written a simple \textsc{Matlab} solver for the Orr--Sommerfeld eigenvalue problem.
The base velocity profile is taken from the interactive boundary layer computations,
and the Orr--Sommerfeld problem is solved indepentently as desired for each value of $x$, $k$, and $t$.
The $y_P$ variable is again truncated at $L_{y_P} = 64$,
by using artificial boundary conditions $\omega(L_{y_P}) = 0$ and $\phi'(L_{y_P}) + k\nu^{1/2} \phi(L_{y_P}) = 0$,
which follow from the reconnection with a potential solution at large $y_P$.

A second order finite differences scheme is used for spatial discretization, written using sparse matrices for efficiency,
which leads to a complex, non-symmetric eigenvalue problem.
The six eigenvalues with largest imaginary part are solved for using the \textsc{Matlab} function ``eigs'',
which relies on implicitly restarted Arnoldi method from \textsc{Arpack}.
As a result, the eigenvalue with the largest imaginary part is readily obtained,
and the unstable wavenumber range can thus be detected and estimated.

\section{Results}
%
%
\subsection{Before detachment}

The behavior of the various solutions well before the Prandtl singularity time
is well understood and we present it only for the sake of completeness.
\begin{figure}
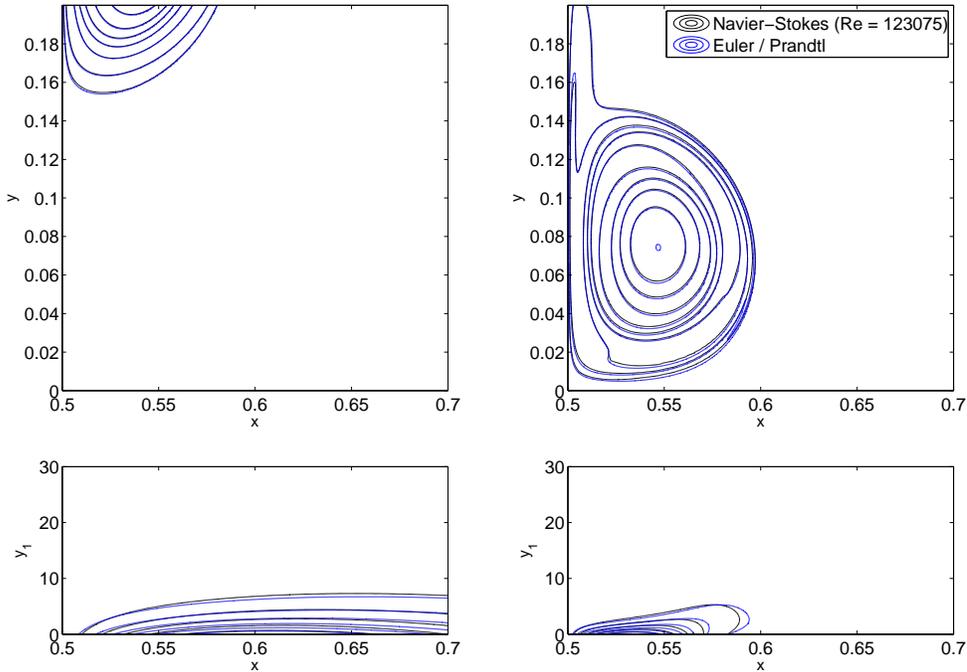

\begin{center}
\includegraphics[width=0.49\columnwidth]{{{FIGURES/bulk_vorticity_t=30}}}
\includegraphics[width=0.49\columnwidth]{{{FIGURES/bulk_vorticity_t=50}}} \\
\includegraphics[width=0.49\columnwidth]{{{FIGURES/boundary_layer_vorticity_t=30}}}
\includegraphics[width=0.49\columnwidth]{{{FIGURES/boundary_layer_vorticity_t=50}}}
\end{center}
\caption{
\label{fig:vorticity_before}
Top: contour lines of vorticity field at $t=30$ (left) and $t=50$ (right), for $\omega = 0.1, 0.2, \ldots, 1$.
Bottom: contour lines of corresponding boundary layer vorticity fields in Prandtl variables,
for $\omega = -\vert \omega\vert _{max}/6, -2\vert \omega\vert _{max}/6, \ldots, -5\vert \omega\vert _{max}/6$, where 
$\vert \omega\vert _{max} = 3.75 \cdot 10^{-4}$ and $\vert \omega\vert _{max} = 8.70 \cdot 10^{-3}$, respectively.
}
\end{figure}
After a rapid relaxation phase, the initial vorticity distribution splits into two counter-propagating dipoles,
each of which shoots towards one of the channel walls.
At this point, the Navier--Stokes vorticity field in the bulk flow remains similar to the 
Euler vorticity field, as shown by comparing their contour lines at $t=30$ in Fig.~\ref{fig:vorticity_before} (top, left).
The Navier--Stokes flow in the Prandtl boundary layer units (bottom, left) is smooth, 
and well approximated by the corresponding solution of the Prandtl equations, shown in blue.
As the dipole approaches the wall, the pressure gradient increases,
causing inward diffusion of vorticity as well as increased vorticity gradients within the boundary layer. 
At $t=50$, we still observe qualitative similarity between the Navier--Stokes flow at high Reynolds number on the one hand, 
and the Euler flow in the bulk with the Prandtl flow in the boundary layer on the other hand (Fig.~\ref{fig:vorticity_before}, right).
However, as expected, the discrepancy between Prandtl and Navier--Stokes flows has increased. 

\begin{figure}
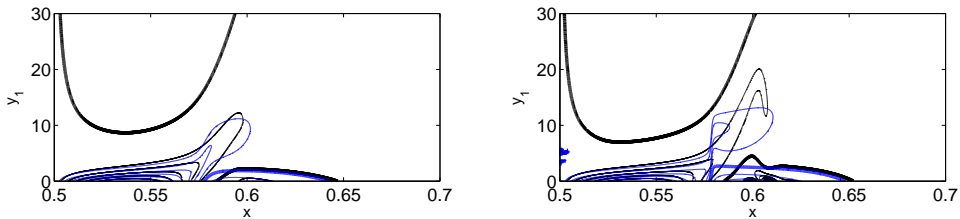

\begin{center}
\includegraphics[width=0.49\columnwidth]{{{FIGURES/boundary_layer_vorticity_t=54}}}
\includegraphics[width=0.49\columnwidth]{{{FIGURES/boundary_layer_vorticity_t=55.3}}}
\end{center}
\caption{
\label{fig:vorticity_before2}
Contour lines of vorticity field in Prandtl variables at $t=54$ (left) and $t=55.3$ (right), 
for $\omega = 5\vert \omega\vert _{max}/6, \ldots, -\vert \omega\vert _{max}/6, 0, -\vert \omega\vert _{max}/6, -2\vert \omega\vert _{max}/6, \ldots, -5\vert \omega\vert _{max}/6$, where 
$\vert \omega\vert _{max} = 1.25 \cdot 10^{-2}$ and $\vert \omega\vert _{max} = 1.30 \cdot 10^{-2}$, respectively.
The $\omega = 0$ contour line is shown in bold.
}
\end{figure}

At $t=54$ (Fig.~\ref{fig:vorticity_before2}, left) a new important feature of the flow is that a region of opposite sign vorticity has appeared within the boundary layer, indicating the build-up of a recirculation bubble along the wall.
This effect is well captured by the Prandtl flow, which overall continues to approach the Navier--Stokes solution pretty well, 
although the discrepancy has again notably increased.

\subsection{Prandtl blow-up}

First signs of a qualitatively different behavior become visible shortly thereafter, 
as shown for example at $t=55.3$ in Fig.~\ref{fig:vorticity_before2} (right).
The contour lines of the Prandtl vorticity have become very concentrated around $x^* = 0.556$, 
indicating the formation of a finite time singularity with precisely the qualitative features predicted by \cite{VanDommelen1990}, 
in particular a blow-up of the wall-normal velocity associated to an infinite acceleration of fluid particles away from the wall.

As the Prandtl solution approaches its singularity time $t^* \simeq 55.6$, 
parallel vorticity gradients increase rapidly, and soon the cut-off parallel wavenumber of the numerical scheme becomes insufficient to resolve it. 


\subsection{Large scale interaction and instability}

According to \cite{Elliott1983}, the Navier--Stokes solution departs from the Prandtl behavior
when the potential flow perturbation due to the presence of the boundary layer
starts to perturb the wall pressure gradient.
\begin{figure}
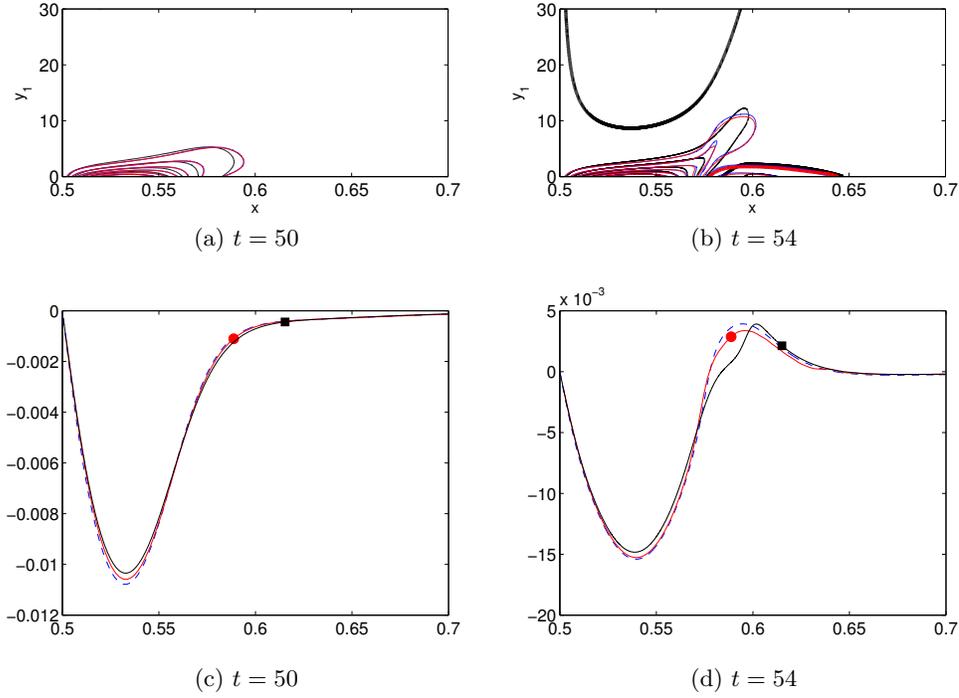

\begin{center} 
\subfloat[$t=50$]{\includegraphics[width=0.49\columnwidth]{{{FIGURES/wet_boundary_layer_vorticity_t=50}}}}
\subfloat[$t=54$]{\includegraphics[width=0.49\columnwidth]{{{FIGURES/wet_boundary_layer_vorticity_t=54}}}} \\
\subfloat[$t=50$]{\includegraphics[width=0.49\columnwidth]{{{FIGURES/wet_vorticity_boundary_t=50}}}}
\subfloat[$t=54$]{\includegraphics[width=0.49\columnwidth]{{{FIGURES/wet_vorticity_boundary_t=54}}}}
\end{center}
\caption{
\label{fig:interactive}
Comparison between Navier--Stokes (black), Prandtl--Euler (blue) and interactive boundary layer (red) models for varying Reynolds numbers,
at $t=50$ (left) and $t=54$ (right).
}
\end{figure}

By comparing the Navier--Stokes, Prandtl and interactive boundary layer models at different Reynolds numbers 
for $t=50$, we observe good agreement between all models (Fig.~\ref{fig:interactive} (a)).
For  $t=54$ (Fig.~\ref{fig:interactive} (b)) it can be noted that the IBL solution has indeed slightly departed away from the 
Prandtl solution, but to a point which fails way short of capturing the full behavior of the NS solution.
This effect, which corresponds in principle to the large scale interaction also described by \cite{Gargano2014},
seems to play only a secondary role in our setting.
More importantly, we observe the growth of an elbow feature in the NS solution,
indicating the start of the growth of a packet of higher-$k$ modes concentrated around $x=0.6$.
\begin{figure}
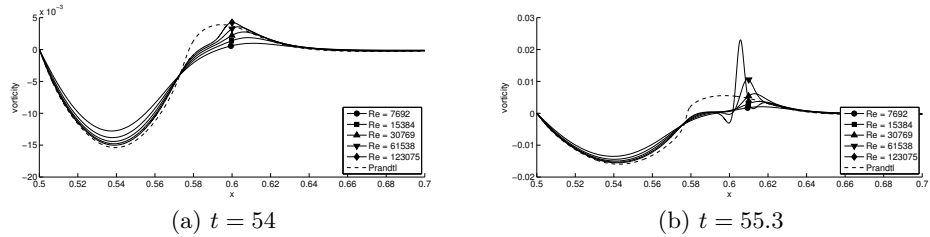

\begin{center}
\subfloat[$t=54$]{\includegraphics[width=0.49\columnwidth]{{{FIGURES/vorticity_boundary_reynolds_t=54}}}}
\subfloat[$t=55.3$]{\includegraphics[width=0.49\columnwidth]{{{FIGURES/vorticity_boundary_reynolds_t=55.3}}}}
\end{center}
\caption{
\label{fig:boundary_vorticity}
Boundary vorticity for varying Reynolds number.
}
\end{figure}
\begin{figure}
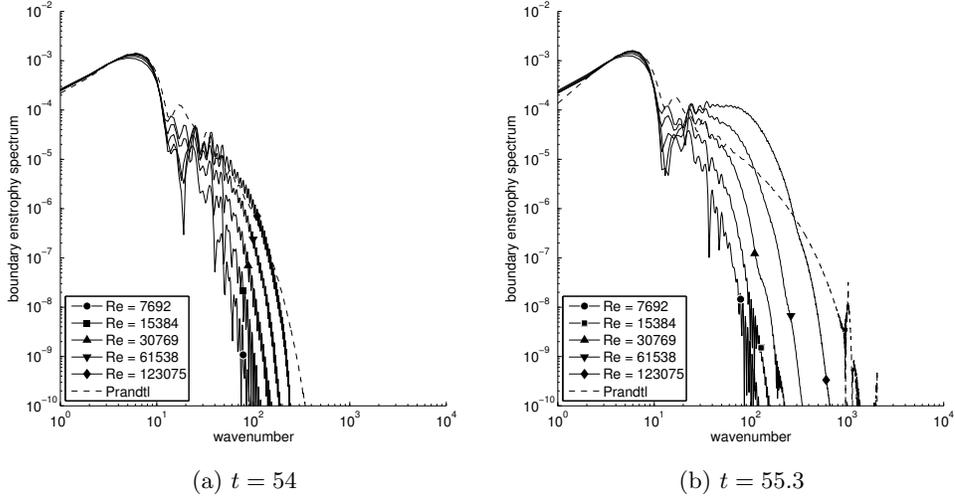

\begin{center}
\subfloat[$t=54$]{\includegraphics[width=0.49\columnwidth]{{{FIGURES/vorticity_boundary_spectra_reynolds_t=54}}}}
\subfloat[$t=55.3$]{\includegraphics[width=0.49\columnwidth]{{{FIGURES/vorticity_boundary_spectra_reynolds_t=55.3}}}}
\end{center}
\caption{
\label{fig:boundary_vorticity_spectra}
Fourier spectra of boundary vorticity for varying Reynolds number.
The spectra have been smoothed to eliminate fast oscillations due to the odd symmetry of the function (see text).
}
\end{figure}

When considering the NSE solution for varying $\Rey$ at the  instants $t=54$ and $t=55.3$ in Fig.~(\ref{fig:boundary_vorticity}),
we observe that for increasing $\Rey$ the elbow structure looks more and more like a wave-packet confined to a well defined interval on the wall.
To understand better the onset of these oscillations, 
it is tempting to consider
one dimensional Fourier transforms of those wall vorticity traces.
Unfortunately the odd symmetry of the function around the dipole axis
gives rise to fast oscillations in the Fourier coefficients
which impair the readability of the spectra.
To get rid of this effect, the spectra are averaged out using the low-pass filter $\exp(-(5.2\vert k \vert )/N)x^{16})$.
The results are shown in Fig.~\ref{fig:boundary_vorticity_spectra}.

The Prandtl solution (dashed curves) develops a $k^{-3/2}$ power law profile at high $k$,
consistent with the build-up of a jump singularity in $\omega$ along the wall.
The NSE solutions spectra all develop a distinctive bump in a wavenumber
range.
Both the width and $k$ location of this bump increase with Reynolds and with time.
Interestingly, for the largest $\Rey$ considered, 
a relatively good separation of scales can be observed at $t=55.3$ between
the low-$k$ features and the high-$k$ wavepacket,
the transition occurring around $k=20$.
This confirms a posteriori the validity of the slow-varying flow approximation
used in deriving the asymptotic stability results of Sec.~\ref{sec:orr-sommerfeld-model}.
Moreover, all solutions have exponentially decaying spectra at sufficiently large values of $k$,
consistent with their analytic regularity being well resolved in the current numerical setting.

\begin{figure}
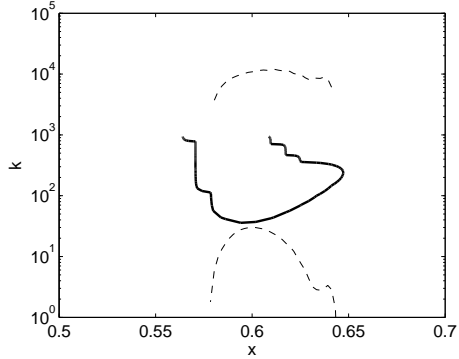

\begin{center}
\includegraphics[width=0.49\columnwidth]{{{FIGURES/fig_unstable_range_t=54}}} 
\end{center}
\caption{
\label{fig:orr-sommerfeld}
Stability boundary subject to the frozen flow approximation, shown as a function of $x$ at $t=54$ and for $\Re = 123075$.
The upper and lower marginal wavenumbers estimated from our asymptotic analysis (Eq.~\ref{eq:critical_k_1} and \ref{eq:critical_k_2}) are
shown in dashed lines.
}
\end{figure}

We now would like to compare the characteristics of these spectra
with the predictions of our analysis based on the Orr--Sommerfeld equations.
The numerical instability range obtained by direct eigenmode analysis (Fig.~\ref{fig:orr-sommerfeld}, bold line)
extends over the interval $x \in [0.555, 0.65]$ on the boundary,
which is in good qualitative agreement with the spatial extent of the oscillations seen in Fig.~\ref{fig:interactive} (b).
In $k$-space, the Orr--Sommerfeld computations predict that the instability should start around $k=30$
for the high $\Rey$ considered, which is in very good agreement as well with the wavenumber at which 
the corresponding spectrum in Fig.~\ref{fig:boundary_vorticity_spectra} (a)
starts to exceed the reference Prandtl solution (shown in dashed lines).
This effect becomes more pronounced at $t=55.3,$ shown in Fig.~\ref{fig:boundary_vorticity_spectra} (b).
Another important point consistent with our scenario is that the stable modes $10 < k < 30$
indeed appear damped in the NSE solutions compared to the Prandtl solution,
a phenomenon which would be very hard to explain using a singularity-type scenario.

Concerning the theoretical prediction for the lower end of the instability range, qualitative agreement is restricted
 to a narrow region around $x=0.6$, whereas the wavenumber is very underestimated 
as soon as a point where $u'_P(0) = 0$ is approached.
Nevertheless, the overall instability region is qualitatively well captured by the criterion $u'_P(0) < 0$.

\subsection{Detachment and production of dissipative structures}

\begin{figure}
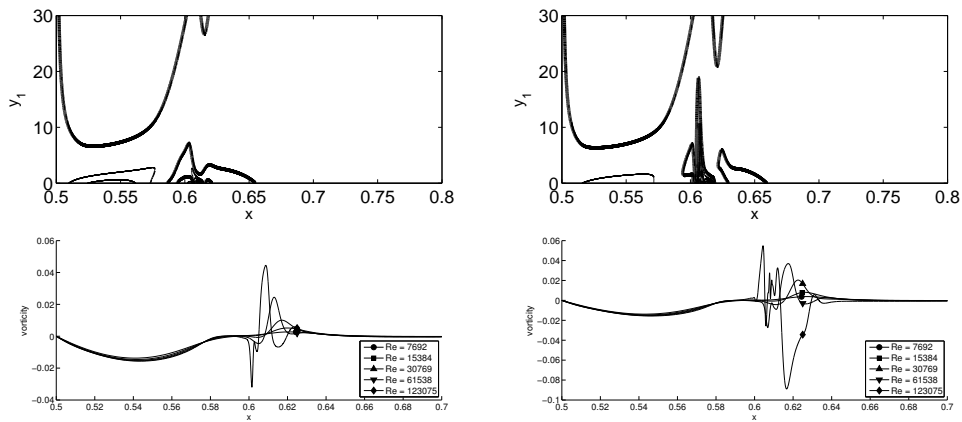

\begin{center}
\includegraphics[width=0.49\columnwidth]{{{FIGURES/boundary_layer_vorticity_t=56}}}
\includegraphics[width=0.49\columnwidth]{{{FIGURES/boundary_layer_vorticity_t=56.9}}} \\
\includegraphics[width=0.49\columnwidth]{{{FIGURES/vorticity_boundary_reynolds_t=56}}}
\includegraphics[width=0.49\columnwidth]{{{FIGURES/vorticity_boundary_reynolds_t=56.95}}}
\end{center}
\caption{
\label{fig:dissipative_structure}
Contour lines of vorticity field in Prandtl variables at $t=56$ (left) and $t=56.9$ (right).
}
\end{figure}

The instability process which we have seen at play above introduces a new vorticity scaling, $\Rey^{1}$, very close to the wall.
This new scaling is difficult to notice at first, because it is hidden behind the preexisting large negative vorticity of the boundary layer.
\begin{figure}
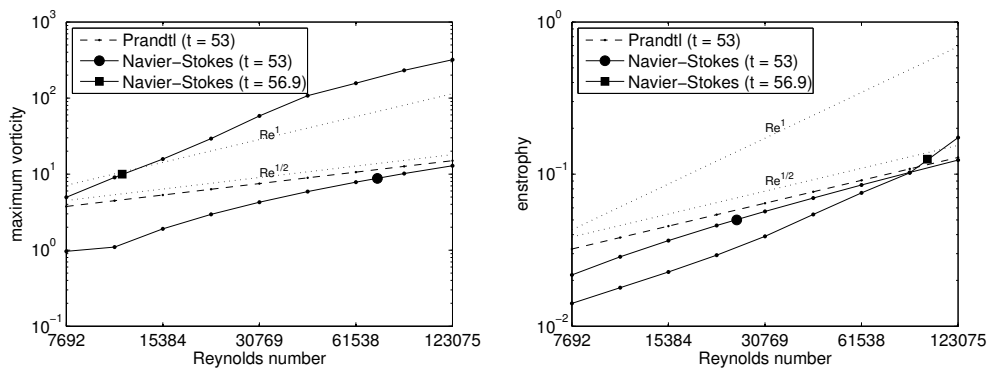

\begin{center}
\includegraphics[width=0.49\columnwidth]{{{FIGURES/maximum_vorticity_reynolds}}}
\includegraphics[width=0.49\columnwidth]{{{FIGURES/enstrophy_reynolds}}}
\end{center}
\caption{
\label{fig:navier_stokes_scalars_scaling}
Reynolds number scalings of maximum vorticity (left) and enstrophy (right).
}
\end{figure}
A simple trick to observe it more easily is to consider only the maximum vorticity of positive sign (Fig.~\ref{fig:navier_stokes_scalars_scaling}, left).
This quantity scales like $\Rey^{1/2}$ at $t=53$, and at $t=56.9$ it has clearly transited to the stronger $\Rey^1$ scaling.
Accordingly, the enstrophy scaling has become dissipative at $t=56.9$, thus indicating the production of a dissipative structure
as predicted by the Kato criterion.
Shortly thereafter, several further extrema with alternating signs successively appear for 
sufficiently high Reynolds number, corresponding to increasingly fine parallel 
scales, as illustrated in Fig.~\ref{fig:dissipative_structure}.
%

\subsection{Later evolution}

\begin{figure}
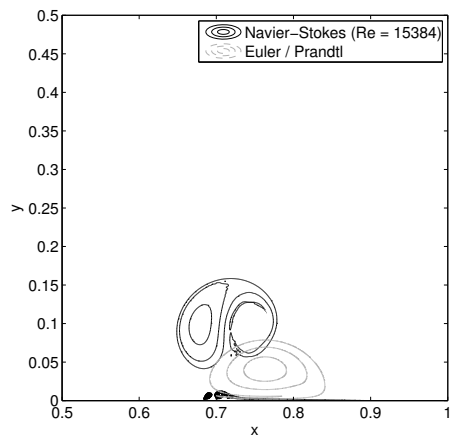

\begin{center}
\includegraphics[width=0.49\columnwidth]{{{FIGURES/bulk_vorticity_t=80.05}}}
\end{center}
\caption{
\label{fig:vorticity_later}
Contour lines of vorticity field at $t=80$.
}
\end{figure}

At much later times, the Euler and Navier--Stokes solutions have 
become entirely different (Fig.~\ref{fig:vorticity_later}).
In the Euler case, the vortices glide along the wall, having paired with their mirror image, and no new vorticity has been generated.
Energy and enstrophy are both conserved.
In the Navier--Stokes case, the detachment process has lead to the formation of two new vortices, 
of much larger amplitude than those of the incoming dipole. 
The activity in the boundary layer remains intense, leading to the ejection of smaller structures.  

\subsection{Convergence checks}

An essential point concerns the control of the discretization error.
Following common practices in numerical fluid dynamics, 
we have taken care of using quite pessimistic scalings to design the wall-parallel and wall-normal grids
in order to resolve the necessary range of scales, 
and as a result we have not observed spurious grid-scale oscillations which would suffice to indicate under-resolution.
%

\begin{figure}
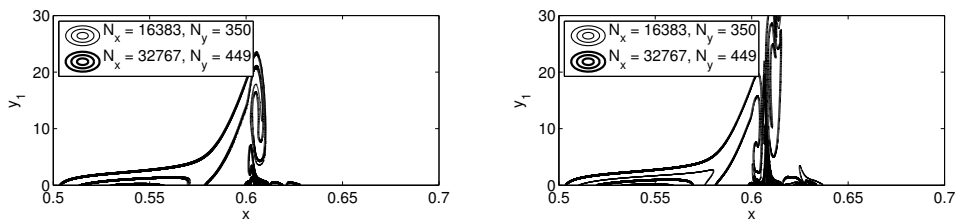

\begin{center}
\includegraphics[width=0.49\columnwidth]{{{FIGURES/boundary_layer_vorticity_ns_t=56_Re=Re_=_123075}}}
\includegraphics[width=0.49\columnwidth]{{{FIGURES/boundary_layer_vorticity_ns_t=56.9_Re=Re_=_123075}}}
\end{center}
\caption{
\label{fig:vorticity_convergence}
Comparison between Navier--Stokes solutions at the same Reynolds number with two numerical resolutions,
in order to check for convergence.
Left: $t=56$. Right: $t=56.9$.
}
\end{figure}

To go one step further, we now consider the contour lines of vorticity during detachment, at $t=56$ and $t=56.9$.
As shown in Fig.~\ref{fig:vorticity_convergence}, the computation used in our analysis and the one obtained with twice the grid spacing
agree well.

\section{Discussion}

There are several features of the numerical solutions which had not been observed in previous work. 
The most striking one is the appearance of the scaling $\Rey^1$ for the vorticity maximum, 
which takes precedence, at the singularity time, 
over the weaker Prandtl scaling $\Rey^{1/2}$.
Even more strikingly, as seen on the graphs of the wall vorticity, this new extremum does not even appear at the same location as the Prandtl singularity. 
This result contradicts sharply the picture of boundary layer detachment as it was described in earlier work, 
as an essentially local process coinciding with the singularity in the Prandtl equations. 
Thanks to the vorticity formulation we have favoured, 
the origin of the non-locality can be traced back to the integral constraints (\ref{eq:integral_bc_old}) on the vorticity field,
which are themselves consequences of the no-slip boundary conditions combined with the incompressibility. 
If higher and higher $k$ modes are excited, as occurs in particular due to the Prandtl singularity formation, the 
reaction of the flow dictated by (1) has no reason to be localized in the $x$ direction. 

A key observation is that in regions with reversed flow near the wall, the width of the unstable wavenumber range scales like $\Rey^{1/2}$,
while the amplitude of vorticity continues to scale as $\Rey^{1/2}$ due to the presence of a Prandtl boundary layer.
Therefore, as soon as the buildup of the Prandtl singularity sufficiently excites those wavenumbers,
their superposition induces a $\Rey^1$ scaling for the amplitude of $\omega$.
In the linear phase, the thickness of the corresponding new wall-normal sublayer scales like $\Rey^{2/3}$,
but as soon as the instability becomes nonlinear, vorticity transport induces excitation of scales as fine as $\Rey^{-1}$,
leading to dissipation.
According to this scenario, the process of detachment is thus intricately linked to the occurrence of dissipation.

Another open question concerns the description of the flow after detachment.
If it is confirmed, the scenario we are proposing indicates that the process of detachment 
and vorticity production by no-slip walls could be modeled by detecting Prandtl singularities
and, when they are about to occur, by introducing nonlinear Rayleigh--Tollmien--Schlichting waves,
followed by roll-up and the injection of a dissipative structure into the bulk flow.
However, an essential point to keep in mind is that the phase of these waves is very sensitive to Reynolds number,
which means that there is little hope of a fully deterministic Reynolds independent description.
This could have important consequences for the modelling of wall-bounded turbulent flows.


The existence of vortical structures in turbulent boundary layers is well established \citep{Robinson1991}.
The local conditions in such flows are therefore not as different from those we have studied
as one might first expect.
According to the logarithmic law of the wall
\begin{equation}\label{eq:law_of_the_wall}
\langle U(y) \rangle \simeq \frac{U_\tau}{K_{\mathrm{karman}}} \log\left(\frac{y U_\tau}{\nu}\right),
\end{equation}
where 
\begin{equation}\label{eq:U_tau_def}
U_\tau = \sqrt{\nu \left\langle\left.\frac{\mathrm{d}U}{\mathrm{d}y}\right\vert_{y=0}\right\rangle}
\end{equation}
is the so-called friction velocity.
This behavior is confirmed by the most recent experiments, with subtle corrections.
An important consequence is that the bulk velocity and $U_\tau$ have the same scaling with $\Rey$ up to a logarithmic factor.
Then, from (\ref{eq:U_tau_def}), one can immediately deduce that $\left.\frac{\mathrm{d}\langle u \rangle}{\mathrm{d}y}\right\vert_{y=0}$
scales like $\Rey^1$ up to a logarithmic factor,
which can be seen as the statistical signature of the existence of a boundary layer of thickness $\Rey^{-1}$ in the neighborhood of the wall.
Hence we see that the log-law, as an experimental result, is consistent in some sense with the existence of a Kato layer, as we have established
in our two-dimensional computations in a much more restricted setting.
This connection can be made, as we just did, in a purely phenomenological way without invoking the Kolmogorov scale and the notion of cascade.
In fact, the essential point is that $U_\tau$ scales with the bulk velocity,
and this scaling was introduced by \citet{Karman1921} precisely to account for the behavior of the drag 
coefficient at high Reynolds number, which was recognized as the essential issue in this time.
From this discussion it appears that our results may help in investigating rigorous foundations to the phenomenological K\'arm\'an theory.

\section*{Acknowledgements}

The authors would like to thank Claude Bardos, David G\'erard-Varet, Helena Nussenzveig Lopes, Milton Lopes Filho, 
and especially Stephen Cowley for important discussions. 
RNVY thanks the Humboldt Foundation for supporting his research by a post-doctoral fellowship.
MF and KS acknowledge support by the French Research Federation for Fusion Studies within the framework of
the European Fusion Development Agreement (EFDA).

\appendix
\section{Estimation of the integral $\boldmath I_c$}\label{sec:I_c_appendix}

In this appendix, we establish the estimate (\ref{eq:I_c_estimate}) for 
the integral $I_c$ defined by (\ref{eq:I_c_def}).
For simplicity we drop the index $P$, and denote generically by $K_R$ real numbers not depending on $c$.
Our guess is that the dominant behavior of $I_c$ when $c\to 0$ is controlled by the behavior of $u$ around its zeros on $[0,+\infty[$.
Hence, let $(y_i)_{1 \leq i \leq m}$ denote the locations of these zeros.
We assume that $u'(y_i) \neq 0$ for all $i$, and that $u$ has an analytic continuation in a complex neighborhood of the real axis,
which ensures that there exists complex neighborhoods $Y_i$ of $y_i$ such that
$u$ is a local holomorphism $u: Y_i \to u(Y_i)$, and $u(Y_i)$ are neighborhoods of $0$.
Assuming that $c$ is sufficiently close to zero so that it lies in the intersection of all the $u(Y_i)$
and it is smaller than the infimum of $u$ outside of the $Y_i$,
the equation $c = u(z)$ has exactly $m$ solutions which we denote $z_i(c)$.
Now letting
\begin{equation}\label{eq:vittali}
f_i(y,c) = \frac{1}{(u(y)-c)^2} - \frac{1}{(u_\infty-c)^2} - \frac{1}{u'(z_i(c))^2(y-z_i(c))^2} + \frac{u''(z_i(c))}{u'(z_i(c))^3(y-z_i(c))},
\end{equation}
it follows from a straightforward Taylor expansion of $u$ around $z_i(c)$ that
$f_i(y,c)$ is bounded on $Y_i$ by a constant independent of $c$, and therefore equiintegrable on $Y_i$.
Therefore, if we define $\xi_0 = 0$, $\xi_{i} = \frac{1}{2}\left(y_{i+1}+y_i\right)$ for $i < m$ and $\xi_m = 2 y_m+1$,
$f_i(y,c)$ is equiintegrable on $[\xi_i,\xi_{i+1}]$,
and by the Vitali convergence theorem, the pointwise limit 
\begin{equation}
f_i(y,0) = \frac{1}{u(y)^2} - \frac{1}{u'(y_i)^2 (y-y_i)^2} + \frac{u''(y_i)}{u'(y_i)^3 (y-y_i) } - \frac{1}{u_\infty^2}
\end{equation}
obtained when $c \to 0$ is integrable,
its integral being the limit of the integrals of $f_i(y,c)$ when $c \to 0$, i.e.
\begin{equation}\label{eq:vitali}
\int_{\xi_{i}}^{\xi_{i+1}} f_i(y,c)\mathrm{d}y \xrightarrow[c \to 0]{} \int_{\xi_{i}}^{\xi_{i+1}} f_i(y,0) \mathrm{d}y
\end{equation}
Now letting
\begin{subequations}
\begin{align}
J_{i,c} = & \int_{\xi_{i}}^{\xi_{i+1}}\frac{\mathrm{d}y}{u'(z_i(c))^2 (y-z_i(c))^2} \\
K_{i,c} = & -\int_{\xi_{i}}^{\xi_{i+1}}\frac{u''(z_i(c))\mathrm{d}y}{u'(z_i(c))^3 (y-z_i(c)) },
\end{align}
\end{subequations}
by construction
\begin{equation}
I_c = -c \sum_{i=0}^{m-1} \int_{\xi_{i}}^{\xi_{i+1}} \left( f_i(y,c)\mathrm{d}y + J_{i,c} + K_{i,c}\right) - c \int_{\xi_m}^\infty \left(\frac{1}{(u(y)-c)^2} - \frac{1}{(u_\infty-c)^2}\right)
\end{equation}
or, by (\ref{eq:vitali}) and the fact that $u$ does not vanish on $[\xi_m,+\infty]$,
\begin{subequations}
\begin{align}
I_c \underset{c\to 0}{=} & -c \sum_{i=0}^{m-1}  \left( \int_{\xi_{i}}^{\xi_{i+1}} f_i(y,0)\mathrm{d}y + J_{i,c} + K_{i,c} \right) - c \int_{\xi_m}^\infty \left(\frac{1}{u(y)^2} - \frac{1}{u_\infty^2}\right) + o(c) \\
\label{eq:I_c_estimate0}
\underset{c\to 0}{=} & -c \sum_{i=0}^{m-1}\left( J_{i,c} + K_{i,c} \right) + c K_R + o(c)
\end{align}
\end{subequations}
so that the dominant behavior of $I_c$ is controlled by $J_{i,c}$ and $K_{i,c}$, which we now proceed to estimate.
Since 
\begin{equation}
J_{i,c}  = -\frac{1}{u'(z_i(c))^2} \left(\frac{1}{\xi_{i+1}-z_{i}(c)} - \frac{1}{\xi_{i}-z_i(c)} \right),
\end{equation}
$J_{i,c}$ converges to a real constant $J_i$ if $i \neq 0$, but diverges for $i=0$.
By Taylor expanding $u$ around $0$, we get that
\begin{equation}
z_0(c) \underset{c\to 0}{=} \frac{c}{u'(0)}  + o(c),
\end{equation}
and therefore
\begin{equation}\label{eq:J_estimate}
J_{0,c}  \underset{c\to 0}{=} -\frac{1}{u'(0) c} + K_R + o(1).
\end{equation}
Now
\begin{align}
K_{i,c} = & - \frac{u''(z_i(c))}{u'(z_i(c))^3} \left( \ln(\xi_{i+1}-z_i(c) ) - \ln(\xi_{i}-z_i(c) ) \right),
\end{align}
where for the complex logarithm we have legitimately taken the principal branch,
since the integration path does not cross the negative real axis.
As before we first assume that $i \neq 0$, in which case
\begin{subequations}
\begin{align}
K_{i,c} \underset{c\to 0}{=} & - \frac{u''(y_i)}{u'(y_i)^3} \left( \ln(\xi_{i+1}-y_i ) - \ln(\xi_{i}-y_i ) \right) + o(1) \\
\label{eq:K_estimate1}
\underset{c\to 0}{=} & \iota \pi \frac{u''(y_i)}{u'(y_i)^3}  + K_R + o(1),
\end{align}
\end{subequations}
whereas we obtain
\begin{equation}
\label{eq:K_estimate2}
K_{0,c} \underset{c\to 0}{=} - \frac{u''(0)}{u'(0)^3} \left( - \ln\left(-\frac{c}{u'(0)} \right) \right) + K_R + o(1)
\end{equation}

Finally, combining (\ref{eq:I_c_estimate0}), (\ref{eq:J_estimate}), (\ref{eq:K_estimate1}) and (\ref{eq:K_estimate2}),
we get the desired estimate of $I_c$:
\begin{equation}\label{eq:I_c}
I_c \underset{c\to 0}{=} \frac{1}{u'(0)} - c \frac{u''(0)}{u'(0)^3} \ln\left(-\frac{c}{u'(0)} \right) -c i \pi \sum_{i=1}^{m-1} \frac{u''(y_i)}{u'(y_i)^3} + c K_R + o(c)
\end{equation}

If $u'(0) = 0$, we cannot apply Vitali's theorem to (\ref{eq:vittali}) for $i=0$ because the second and third term diverge when $c \to 0$.
\cite{Hughes1965} computed the asymptotic expansion of $I_c$ for a special form of $u$ with $u'(0)=0$, 
but we rederive it using a different method for completeness.
Letting $v = \sqrt{u}$,
we may write
\begin{equation}
(u-c)^2 = (v-\sqrt{c})(v+\sqrt{c}),
\end{equation}
with 
\begin{equation}
v'(z)^2 \xrightarrow[z \to 0]{} \frac{u''(0)}{2}, \quad v''(z) \xrightarrow[z \to 0]{} \pm \sqrt{\frac{2}{u''(0)}}u'''(0),
\end{equation}
so that the above results can be applied to $v$ with $z^+$ and $z^-$ defined by $v(z^+(c)) = \sqrt{c}$, $v(z^-(c)) = \sqrt{c}$,
leading to
\begin{equation}\label{eq:vittali2}
g^\pm(y,c) = \frac{1}{(v(y) \pm \sqrt{c})^2}  - \frac{1}{v'(z^\pm(c))^2(y-z^\pm(c))^2} + \frac{v''(z^\pm(c))}{v'(z^\pm(c)^3(y-z^\pm(c))}.
\end{equation}
Since the functions $g^\pm$ are bounded by a constant independent of $c$, 
we can now safely apply the Vitali theorem to the product $g^+ g^-$.
Due to the order of the different terms, it is sufficient to keep only one of them
\begin{equation}
I_c \underset{c \to 0}{\sim} -c\int_0^\infty \frac{1}{v'(z^+(c))^2 v'(z^-(c))^2(y-z^+(c))^2(y-z^-(c))^2},
\end{equation}
and after computing the residuals we finally obtain
\begin{equation}\label{eq:I_c_up_0}
I_c \underset{c\to 0}{=} (8u''(0)c)^{-1/2}\iota\pi + o(c^{-1/2})
\end{equation}

\section{Validation of the solvers}\label{sec:validation_appendix}

Although the discretization methods used for this paper are relatively classical, 
the way the boundary conditions are imposed is new 
and it was thus necessary to conduct validation runs which are reported here.

\subsection{Navier--Stokes solver}
	
As test case for the Navier--Stokes solver, the setup designed by \cite{Kramer2007} was considered.
Contrary to the quadrupole setup which has been studied in the body of the present paper,
the dipole is not symmetric with respect to the channel centerline.
The full span of the channel and the walls on both sides therefore needs to be taken into account.
Two runs with $\Rey=650$ and $\Rey=2500$ were performed, respectively with $512 \times 255$ and $1024 \times 511$ uniformly distributed grid points.

\begin{table}
\begin{center}
\begin{tabular}{cccccccc}
\hline\hline
	   &     &  \multicolumn{3}{c}{Current results} & \multicolumn{3}{c}{\cite{Kramer2007}} \\
\cmidrule(l){3-5}
\cmidrule(l){6-8}
$\Rey$ & $t$ & $x_d$ & $y_d$ & $\omega_d$ & $x_d$ & $y_d$ & $\omega_d$ \\
\hline
 	& $ 0.0 $ 	& $ 0.0977 $ 	& $ 1.0000 $	& $ 316.70 $ 	& $ 0.1000 $ 	& $ 1.0000 $ 	& $ 316.84 $	\\
$625$ 	& $ 0.6 $ 	& $ 0.1680 $ 	& $ 0.1836 $	& $ 158.51 $ 	& $ 0.1656 $ 	& $ 0.1827 $ 	& $ 158.70 $	\\
$625$ 	& $ 1.0 $ 	& $ 0.2578 $ 	& $ 0.1953 $	& $ 102.53 $ 	& $ 0.2543 $ 	& $ 0.1949 $ 	& $ 102.64 $	\\
$2500$ 	& $ 0.6 $ 	& $ 0.1797 $ 	& $ 0.1016 $	& $ 261.83 $ 	& $ 0.1654 $ 	& $ 0.1042 $ 	& $ 261.83 $	\\
$2500$ 	& $ 1.0 $ 	& $ 0.1562 $ 	& $ 0.1445 $	& $ 231.22 $ 	& $ 0.2185 $ 	& $ 0.1738 $ 	& $ 231.40 $	\\
\hline 
\hline
\end{tabular}
\end{center}
\caption{
\label{tab:validation_nse}
Location $(x_d,y_d)$ and value $(\omega_d)$ of maximum vorticity for various times and Reynolds numbers, 
compared to the data in Table~III of \cite{Kramer2007}, for the dipole-wall test case.}
\end{table}
In order to make a quantitative comparison, the same procedure used by \cite{Kramer2007} is repeated here,
namely to compare the location and amplitude of the main vortex core at several instants.
The data are presented in Table~\ref{tab:validation_nse}.
Note that at $t=0$, there is a minor misstake in the reference data, since $x_d=0.1$ corresponds by construction to the location of the vorticity maximum for an isolated monopole, whereas the maximum of the dipole is slightly shifted due to interaction with the opposite sign vortex.
The results are otherwise in good agreement.

\subsection{Prandtl--Euler solver}

For the Prandtl solver, the classical impulsively started cylinder studied by VDS in the Lagrangian framework is employed as a test case.
It corresponds to a constant boundary pressure gradient given by
\begin{equation}
\partial_x p_E(x,0,t) = -\sin(x)\cos(x),
\end{equation}
and an initial vorticity which is a Dirac distribution
\begin{equation}
\omega_P(x,y,0) = -\delta_0(y)\sin(x)
\end{equation}

For spatial discretization, $1023$ grid points are considered on the interval $[0,\pi]$ in the $x$ direction,
taking advantage of the odd symmetry of the solution,
and $513$ grid points with $L_y=32$ in the $y$ direction.
The initial Dirac distribution is approximated by letting
\begin{equation}
\omega_P(x_i,0,0) = -\frac{\sin(x_i)}{\Delta x}, \quad \omega_P(x_i,y,0) = 0 \mathrm{\ if\ } y > 0.
\end{equation}
The results are then compared with Fig.~10 and Table~II of VDS.
\begin{figure}
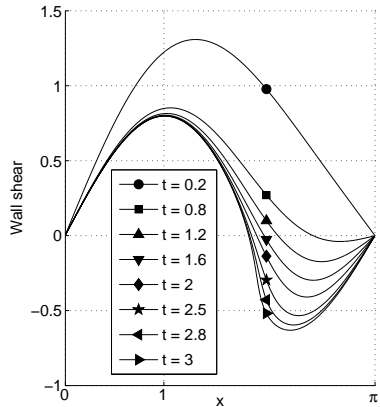

\begin{center}
\includegraphics[width=0.4\columnwidth]{{{FIGURES/prandtl_validation}}}
\end{center}
\caption{
\label{fig:prandtl_validation}
Wall shear stress at different instants for the impulsively started cylinder test case.
}
\end{figure}
Fig.~\ref{fig:prandtl_validation} shows the wall shear stress for different time instants.
It is in very good qualitative agreement with Fig.~10 of VDS, except maybe at very short times
which is to be expected given the singular initial condition.
Additionally, our estimate for the quantity $F''_w$ at $t=3$ is $1.1061$, which is in good agreement with the value $1.1122$ found by VDS
at their much lower resolution (in doing this comparison we have assumed that the undefined quantity $T$ appearing in Table~II of VDS corresponds to $t/2$).

\subsection{Orr--Sommerfeld solver}

To validate the Matlab code used to compute Orr--Sommerfeld eigenvalues,
we use the Blasius boundary layer as a test case.
\begin{figure}
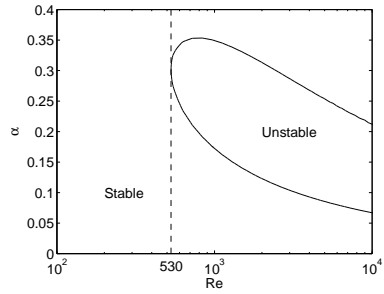

\begin{center}
\includegraphics[width=0.4\columnwidth]{{{FIGURES/orr_sommerfeld_validation}}}
\end{center}
\caption{
\label{fig:orr_sommerfeld_validation}
Unstable region for the Blasius boundary layer.
}
\end{figure}
The obtained marginal stability curve (Fig.~\ref{fig:orr_sommerfeld_validation}) 
is in good agreement with Fig.16.11 provided by \cite{Schlichting1979}.

\newpage
\section{Overall flow evolution: comparison of Navier--Stokes and Euler/Prandtl flows} 
\begin{figure}
\begin{center}
\includegraphics[width=1.0\columnwidth]{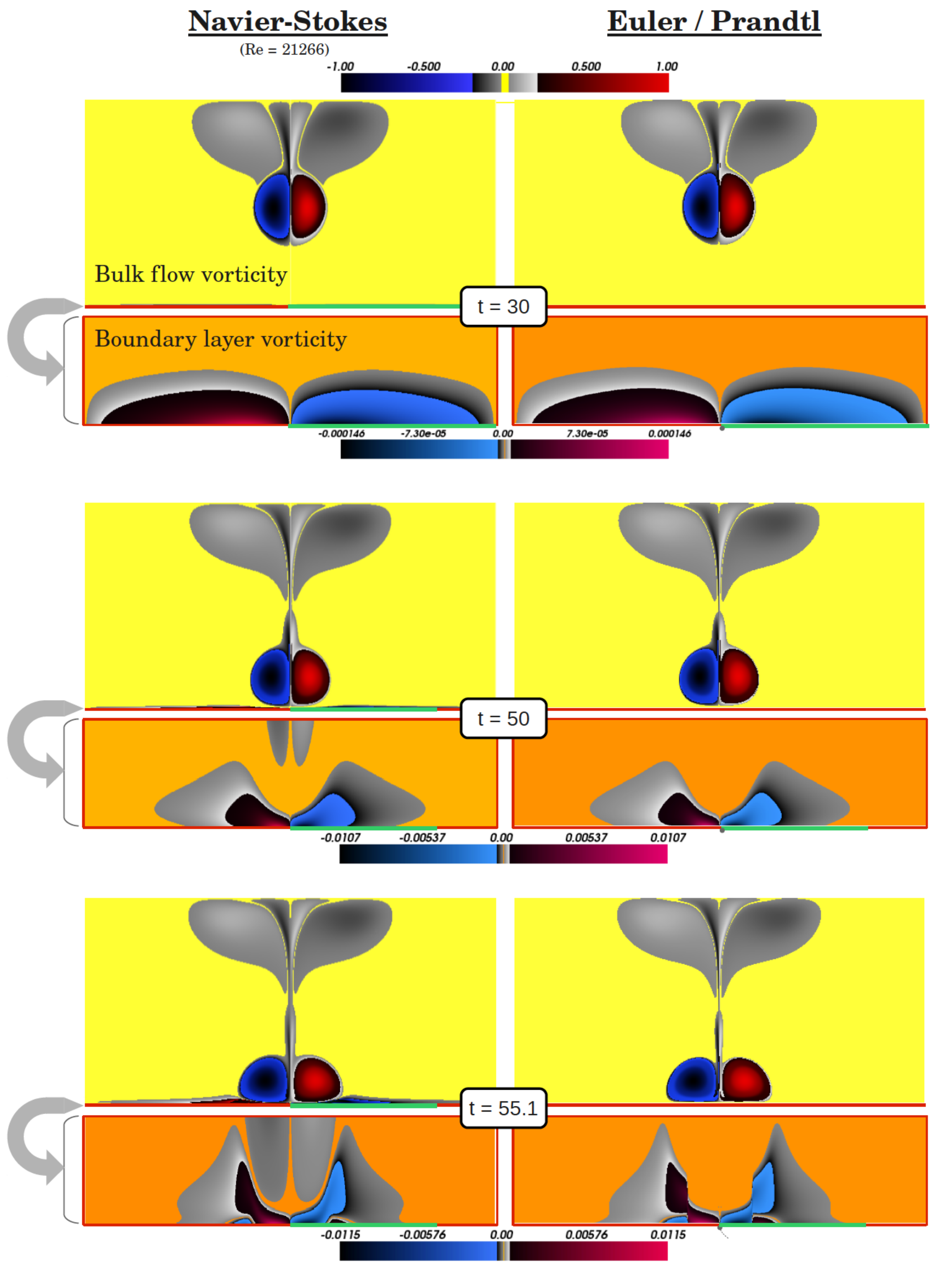}
\end{center}
\caption{
\label{fig:testposter1}   
Vorticity fields for the Navier--Stokes solution (left) and the Euler solution (right) in the bulk flow
at $t=30, 50$ and $55.1$. The corresponding vorticity fields in the boundary layer for the Navier--Stokes (left) and the Prandtl solutions (right) 
are given directly below.}
\end{figure}

The dipole first shoots towards the lower channel wall. The Navier--Stokes vorticity field in the bulk (Fig.~\ref{fig:testposter1} top, left) remains very close to 
the Euler vorticity field (Fig.~\ref{fig:testposter1} top, right).
Plotting the Navier-Stokes flow in the Prandtl boundary layer units
(Fig.~\ref{fig:testposter1} top, left) reveals that it is smooth, and very well approximated
by the solution of the Prandtl equations (Fig.~\ref{fig:testposter1} top, right).
The vorticity along the boundary (Fig.~\ref{fig:testposter3}, top, left) converges to the Prandtl values as the Reynolds number increases.

As the dipole approaches the wall (Fig.~\ref{fig:testposter1}, middle), the pressure gradient along the wall becomes more intense and steeper, which causes a
strong inward diffusion of vorticity at the wall, as well as increased vorticity gradients within the boundary layer.
At time $t=50$, we still observe convergence of the Navier--Stokes solution at high Reynolds number towards the Euler flow in the bulk
and towards the Prandtl flow in the boundary layer. However, looking at the boundary vorticity (Fig.~\ref{fig:testposter3}, top, right) reveals a larger
difference between Prandtl and Navier--Stokes flows than at $t=30$.

As the Prandtl solution approaches its singularity time $t^* \approx 55.6$ (Fig.~\ref{fig:testposter1}, bottom),
parallel vorticity gradients increase rapidly, and soon the cut-off parallel wavenumber of the numerical scheme becomes insufficient
to resolve it. The convergence of the Navier--Stokes boundary vorticity is lost over a wide interval in $x$, and the vorticity around $x=0.61$ adopts a
stronger scaling with Reynolds (Fig.~\ref{fig:testposter3}, bottom, left).

\begin{figure}
\begin{center}
\includegraphics[width=1.0\columnwidth]{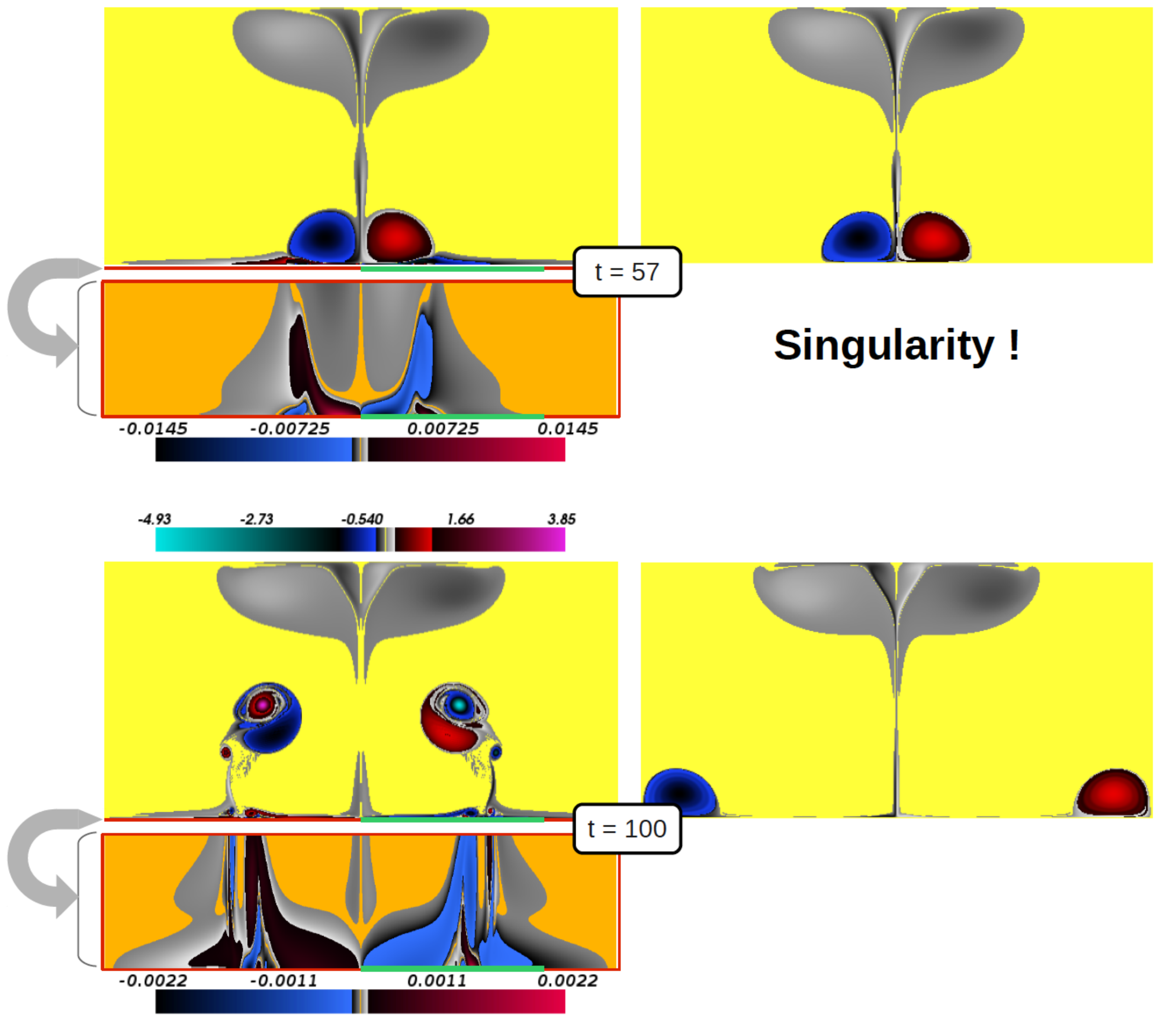}
\end{center}
\caption{
\label{fig:testposter2}
Continuation of Fig.~\ref{fig:testposter1} at time instants $t= 57$ and $100$. The Prandtl solution is singular and thus not shown.
}
\end{figure}

\begin{figure}
\begin{center}
\includegraphics[width=.45\columnwidth]{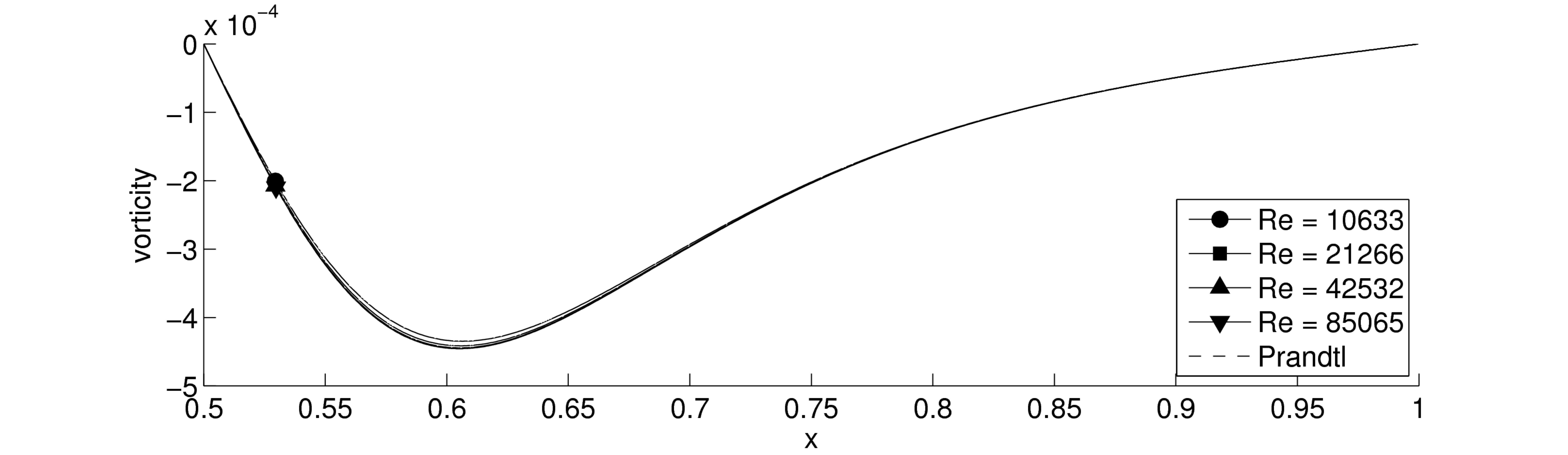}
\includegraphics[width=.45\columnwidth]{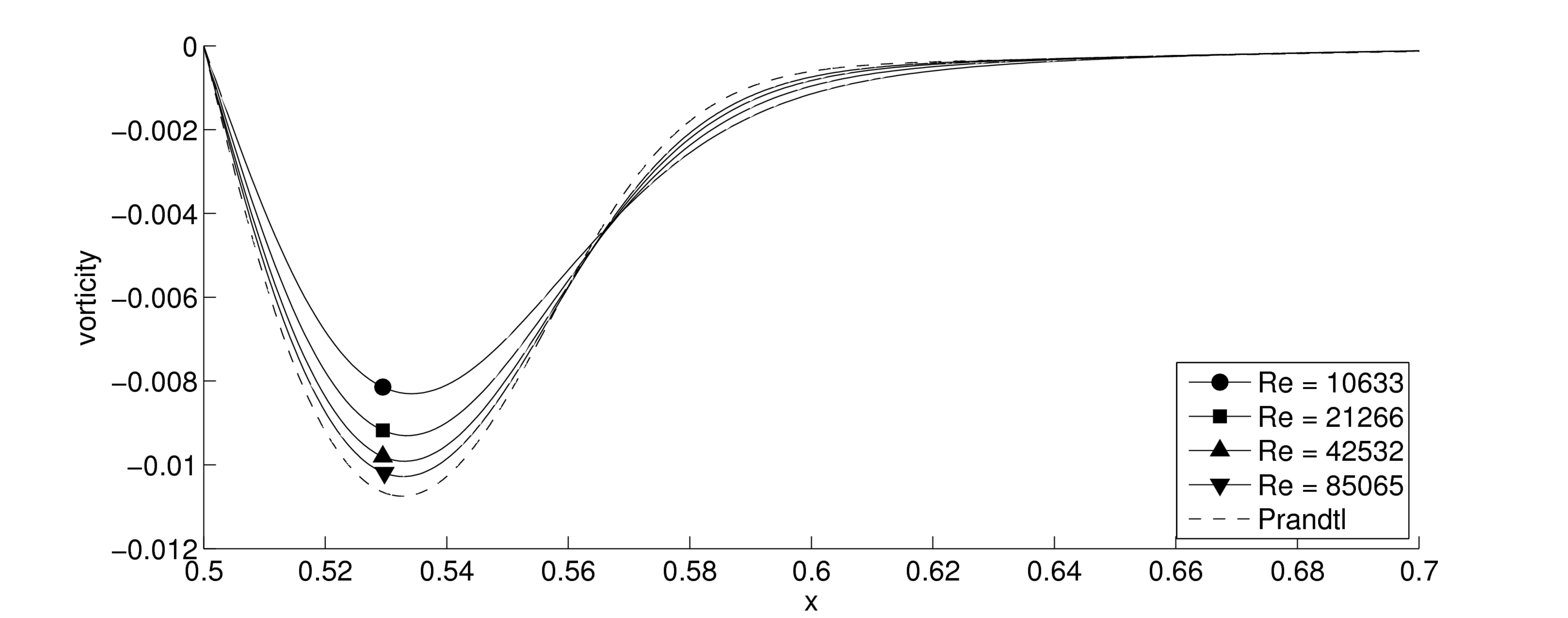}\\ \vspace{0.2cm}
\includegraphics[width=.45\columnwidth]{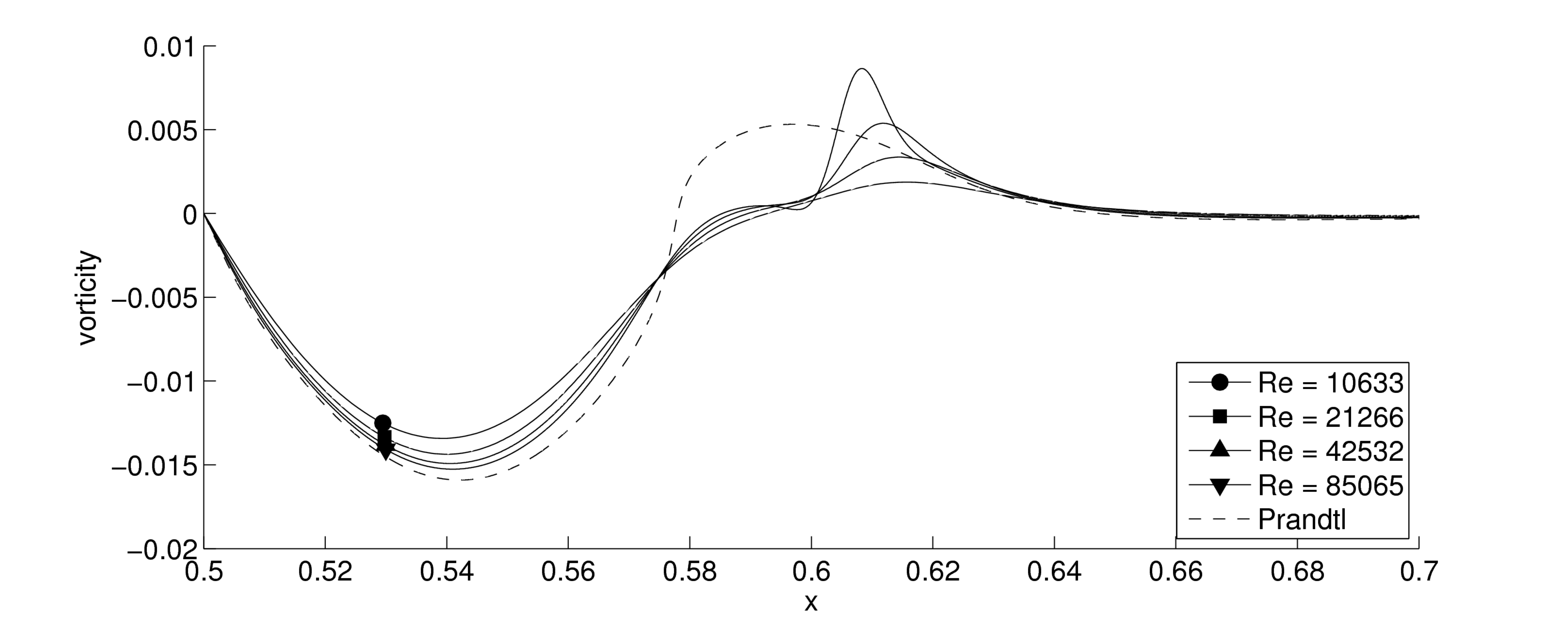}
\includegraphics[width=.45\columnwidth]{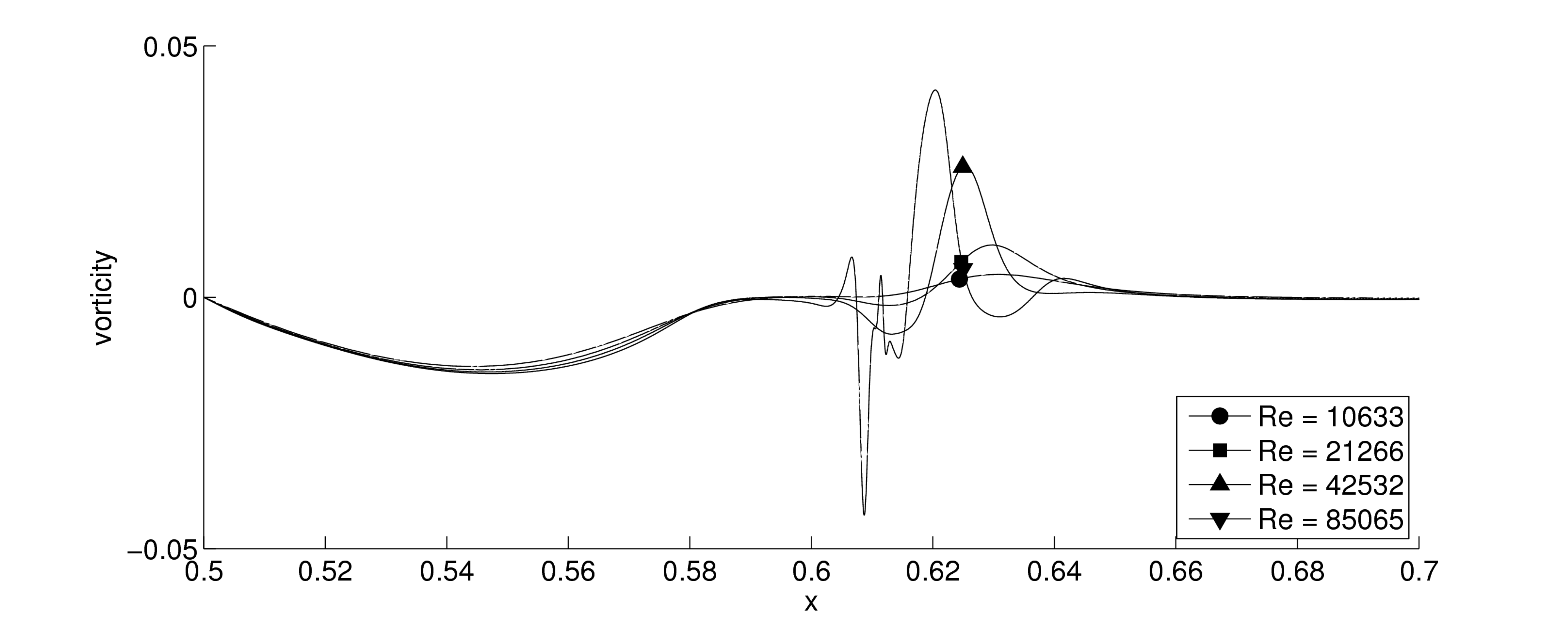}
\end{center}
\caption{
\label{fig:testposter3}
Vorticity along the boundary at $t=30, 50, 55.1$ and $57$.
}
\end{figure}

After the singularity, at $t=57$ (Fig.~\ref{fig:testposter2}, top) oscillations in the vorticity appear, while
 the bulk flow still looks similar for Navier--Stokes and Euler.
Following the new vorticity extremum which has appeared at the boundary, a cascade of extrema with opposite signs appear (for
sufficiently high Reynolds number), exciting increasingly fine scales parallel to the wall (Fig.~\ref{fig:testposter3}, bottom, right).

At much later times, $t=100$ (Fig.~\ref{fig:testposter2}, bottom), the Euler and Navier--Stokes solutions have become completely different.
In the Euler case, the vortices glide along the wall, having paired with their mirror image, no new vorticity has been generated and the energy is conserved.In the Navier--Stokes case, the detachment process has lead to the formation of two new vortices (shown in cyan and
magenta in the Fig.~\ref{fig:testposter2}, bottom,left) of much larger amplitude than those of the incoming dipole. The activity in the boundary layer remains
intense, leading to the ejection of smaller structures.


\bibliography{biblio}
\bibliographystyle{jfm}
\end{document}